\documentclass[twocolumn,preprintnumbers,amsmath,amssymb]{revtex4-1} 

\usepackage{graphicx}
\usepackage{dcolumn}
\usepackage{bm}

\newcommand*{\rbT}[1]{\raisebox{0.8ex}[-2.2ex]{#1}} 
\newcommand*{\rbL}[1]{\raisebox{1.8ex}[-1.2ex]{#1}} 

\begin{document}


\title[Analytics of Zinc-Blende Nanocrystals: Size, Shape and Surface Orientation]{Number Series of Atoms, Interatomic Bonds and Interface Bonds Defining Zinc-Blende Nanocrystals as Function of Size, Shape and Surface Orientation: Analytic Tools to Interpret Solid State Spectroscopy Data}


\author{Dirk K{\"o}nig}
\email{dirk.koenig@unsw.edu.au}
\affiliation{Integrated Materials Design Centre (IMDC) and School of Photovoltaic and Renewable Energy Engineering (SPREE),\\ University of New South Wales, Sydney, Australia}
\altaffiliation[Also at ]{Laboratory of Nanotechnology, Dept. of Microsystems Engineering (IMTEK), University of Freiburg, Germany.}


\begin{abstract}
Semiconductor nanocrystals (NCs) experience stress and charge transfer by embedding materials or ligands and impurity atoms. In return, the environment of NCs experiences a NC stress response which may lead to matrix deformation and propagated strain. Up to now, there is no universal gauge to evaluate the stress impact on NCs and their response as a function of NC size $d_{\mathrm{NC}}$. I deduce geometrical number series as analytical tools to obtain the number of NC atoms $N_{\mathrm{NC}}(d_{\mathrm{NC}}[i])$, bonds between NC atoms $N\mathrm{_{bnd}}(d_{\mathrm{NC}}[i])$ and interface bonds $N\mathrm{_{IF}}(d_{\mathrm{NC}}[i])$ for seven high symmetry zinc-blende (zb) NCs with low-index faceting: \{001\} cubes, \{111\} octahedra, \{110\} dodecahedra, \{001\}-\{111\} pyramids, \{111\} tetrahedra, \{111\}-\{001\} quatrodecahedra and \{001\}-\{111\} quadrodecahedra. 
The fundamental insights into NC structures revealed here allow for major advancements in data interpretation and understanding of zb- and diamond-lattice based nanomaterials. The analytical number series can serve as a standard procedure for stress evaluation in solid state spectroscopy due to their deterministic nature, easy use and general applicability over a wide range of spectroscopy methods as well as NC sizes, forms and materials.
\end{abstract}

\maketitle

\section{\label{intro}Introduction}
It is well known that the electronic structure and optical response of NCs is a function of mechanical stress in the form of lattice strain. Mechanical stress is routinely measured by Raman- and Fourier-Transformation InfraRed (FT-IR) spectroscopy which probe the phononic spectra of NCs \cite{Jac66,Ana70,Nak81,Boy82,Boy87}. Such spectra are very sensitive to changes of stress induced by  lattice pressure which is a function of the material via Young's modulus \cite{Ell98}. Changes in compressive or expansive stress were shown to modify the optical response of NCs by fluorescence \cite{Smi09} or photoluminescence (PL) \cite{Cli12,Lau14}. Stranski-Krastanov growth \cite{StraKras38,Bau58} of NCs in epitaxial films depends critically on balanced stress to avoid stacking faults which deteriotate electronic NC properties \cite{Osh06,Osh08,Pope08,Bail09}. Attempts to place phosphorus atoms as donors onto lattice sites in free-standing Si NCs were shown to fail increasingly with shrinking NC diameter \cite{Steg08a,Steg09}, revealing a transition region from low to virtually zero doping of 20 to 10 nm. These findings were confirmed recently in theory and experiment for SiO$_2$-embedded Si NCs \cite{Koe15a,Gnas14}. Several research groups have shown that self-purification, a Si NC-internal build-up of stress counteracting external stress due to dopant incorporation, causes impurity doping to fail \cite{Dalp06,Dalp08,Chan08,Ossi05}. 

With a universal gauge for stress such phenomena could be scaled as a function of NC size $d_{\mathrm{NC}}$. Principal parameters are the number of atoms forming the zb-NC $N_{\mathrm{NC}}(d_{\mathrm{NC}})$, the bonds between such atoms $N\mathrm{_{bnd}}(d_{\mathrm{NC}})$ and the number of bonds $N\mathrm{_{IF}}(d_{\mathrm{NC}})$ terminating the NC interface. The ratio $N\mathrm{_{bnd}}(d_{\mathrm{NC}}[i])/N_{\mathrm{NC}}(d_{\mathrm{NC}}[i])$ yields the bonds per atom within zb-NCs as a gauge for the response to external stress, while the ratio $N\mathrm{_{IF}}(d_{\mathrm{NC}}[i])/N\mathrm{_{bnd}}(d_{\mathrm{NC}}[i])$ describes the ability of embedding materials to exert stress onto NCs. The impact of a highly polar surface termination on the zb-NC electronic structure is assessed by  $N\mathrm{_{IF}}(d_{\mathrm{NC}}[i])/N_{\mathrm{NC}}(d_{\mathrm{NC}}[i])$ which provides a gauge to interface charge transfer \cite{Koe14}.

\begin{figure*}[t!]
\begin{center}
\includegraphics[totalheight=0.1114\textheight]{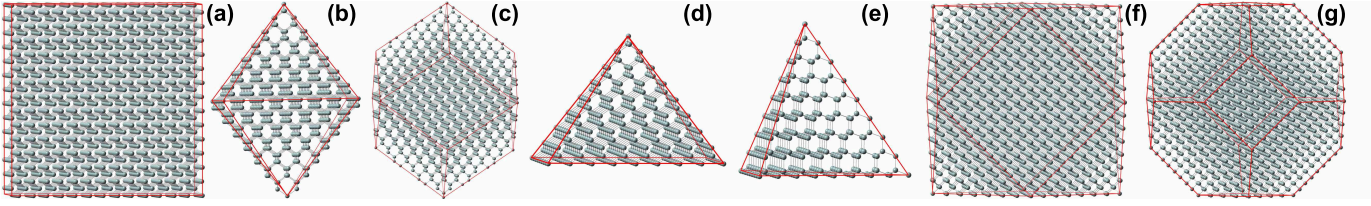}
\end{center}
\caption{{\bf Diamond (zinc blende) lattice NCs considered in this work.} Cubic NCs with exclusive \{001\} surfaces (a), octahedral NCs with exclusive \{111\} surfaces (b), dodecahedral NCs with exclusive \{110\} surfaces (c), pyramidal NCs with \{001\} base and \{111\} side surfaces (d), tetahedral NCs with exclusive \{111\} surfaces (e), quatrodecahedral NCs featuring \{001\}- and \{111\}-surfaces, with dominant \{001\} (f) and \{111\} (g) faceting.}
\label{fig01}
\end{figure*}

While we illustrate our findings on Si NCs (diamond lattice), analytical number series introduced below are also valid for zb-NCs due to straightforward symmetry arguments. Thus, analytical descriptions below cover III-V and II-VI compounds with zb symmetry in addition to Si, SiC, SiGe and Ge. Figure \ref{fig01} shows the regular NC shapes investigated: cubic (\{001\} faceting), octahedral (\{111\} faceting) and dodecahedral (\{110\} faceting); pyramidal (\{001\} base, \{111\} side factes) and tetragonal (\{111\} faceting); 111-quatrodecahedral (dominant \{111\} faceting plus \{001\} faceting) and 001-quatrodecahedral (dominant \{001\} faceting plus \{111\} faceting).

Due to the complexity of symmetry arguments, zb-NCs with higher index faceting are beyond the scope of this work. For Si, the \{111\} (\{001\}) facets have the lowest (second-lowest) experimental values of surface free energy, \emph{cf.} table \ref{tab-00}. The occurence of \{110\} facets on zb-NCs is therefore less likely as compared to those with \{001\} and in particular \{111\} orientation. Facetes with \{111\} orientation have the lowest surface bond density for all facets up to \{433\} orientation \cite{Hes93}. These findings also hold for other diamond- and zb-NCs due to symmetry arguments.  

The number series follow a run index $i$ which relates indirectly to $d_{\mathrm{NC}}[i]$ via $N_{\mathrm{NC}}[i]$. This connection is made by the atomic volume $V_{\mathrm{atom}}$, yielding the NC volume as $V_{\mathrm{NC}} = N_{\mathrm{NC}}[i]\times V_{\mathrm{atom}}$. Since NCs are often described as spherical, we use $d_{\mathrm{NC}}$ for spherical NCs which allows to compare different NC shapes as function of $d\mathrm{_{NC}}$:
\begin{eqnarray}\label{eqn-01}
  d\mathrm{_{NC}}[i]&=&\sqrt[3]{\frac{6}{\pi}N\mathrm{_{NC}}[i]\times V_{\mathrm{atom}}}
\end{eqnarray}  
\begin{table}[t!]
\caption{Bond densities and free energies per square for low index Si facets. Bond density values taken from \cite{Hes93}, experimental surface energy values taken from \cite{Eagl93}.} \label{tab-00}\vspace*{0.1cm}
\renewcommand{\baselinestretch}{1.2}\small\normalsize
  \begin{tabular}{c|c|c}
    \hline
    facet&surface bond&surface free\\ 
    orientation&density [cm$^{-2}$]&energy [Jm$^{-2}$]\\ \hline 
  \{001\}&$1.36\times 10^{15}$&1.36\\ 
  \{110\}&$0.96\times 10^{15}$&1.43\\ 
  \{111\}&$0.78\times 10^{15}$&1.23\\ 
  \end{tabular}
\renewcommand{\baselinestretch}{1.0}\small\normalsize
\end{table}

\section{\label{NumSeries}Analytical numbers series of nanocrystal types}
The derivation of number series can be lengthy and complex. Therefore, only key equations presenting final results are shown in {\bf bold print} for all zb-NC types. The general algorithm is briefly explained for \{110\}-faceted dodecahedral NCs as an example. For brevity, we only list final results for all other NC types.
\subsection{\{001\} Cubic Zinc-Blende Nanocrystals}
\begin{figure*}[t!]
\begin{center}
\hspace*{-0.19cm}
\mbox{\scalebox{0.0883}{\includegraphics{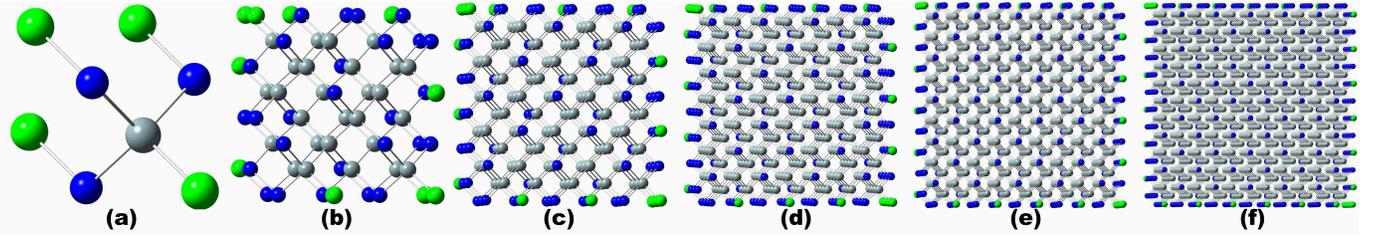}}}
\end{center}
\caption{{\bf Cubic NCs with exclusive \{001\} facets described by equations \ref{eqn-02} to \ref{eqn-04}.} Grey atoms have all bonds in NC. Blue atoms have two interface bonds and green atoms have three interface bonds. Graphs (a) to (f) show NCs for $i=\mbox{1,\,2,\,3,\,4,\,5}$ and 6, respectively, \emph{cf.} table \ref{tab2}.}\label{fig02}
\end{figure*}

Nearly cubic zb-NCs are mainly encountered in nano- and micro-crystalline thin films as found in Si solar cells or ultra-large scale integration (ULSI) devices. The number of atoms forming the NC is given by the product of the atoms of the face-centered-cubic (fcc) unit cell with the cubic increase per index $i$:
\begin{equation}\label{eqn-02}
\mathbf{
N\mathrm{\mathbf{_{NC}^{cube}}}[i]=8\, i^3,\,\forall  i \geq 1\,. }
\end{equation}
The number of bonds between NC atoms is given by  
\begin{equation}\label{eqn-03}
\mathbf{
N\mathrm{\mathbf{_{bnd}^{cube}}}[i]=i\big[(4i-2)(4i-1)+1\big],\,\forall  i \geq 1\,. }
\end{equation}
The surface bonds of the NC are described by
\begin{equation}\label{eqn-04}
\mathbf{
N\mathrm{\mathbf{_{IF}^{cube}}}[i]=6\left[\left(2i-1\right)^2+3i-1\right],\,\forall i \geq 1\,. }
\end{equation}
These number series can be verified in figure \ref{fig02} where atoms are color-coded in accord with their bond configuration.
\subsection{\label{Octa}\{111\} Octahedral Zinc-Blende Nanocrystals}
Exclusively \{111\}-faceted Si NCs correspond to the minimum Si surface energy in experiment \cite{Eagl93} and were proven to exist up to a size of ca. 30 {\AA} \cite{Gode08}. 
\begin{figure*}[t!]
\begin{center}
\mbox{\scalebox{0.0877}{\includegraphics{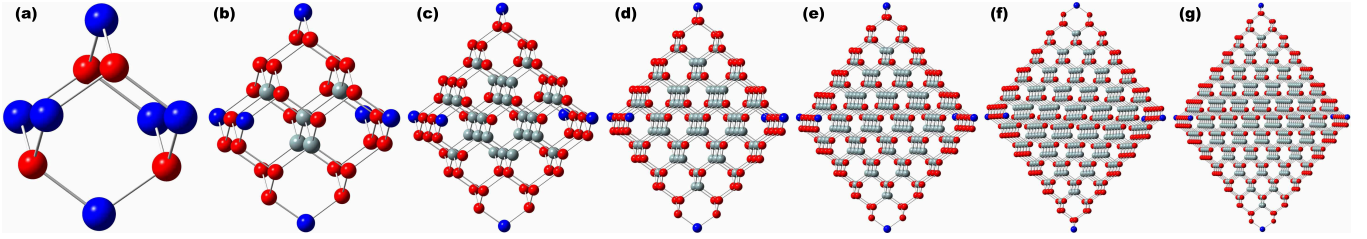}}}
\end{center}
\caption{{\bf Octahedral NCs with exclusive \{111\} facets described by equations \ref{eqn-05} to \ref{eqn-07}.} Grey atoms have all bonds in NC. Red atoms have one interface bond and blue atoms have two interface bonds. Graphs (a) to (f) show NCs for $i=\mbox{1,\,2,\,3,\,4,\,5,\,6}$ and 7, respectively, \emph{cf.} table \ref{tab2}.}
\label{fig03}
\end{figure*}

For the number of atoms forming the NC we obtain 
\begin{equation}\label{eqn-05}
\mathbf{
N\mathrm{\mathbf{_{NC}^{octa}}}[i]=\frac{1}{3}(i+1)\left[4(i+1)^2-1\right],\,\forall i \geq 0\,. }
\end{equation}
The number of bonds between NC atoms is given by 
\begin{equation}\label{eqn-06}
\mathbf{
N\mathrm{\mathbf{_{bnd}^{octa}}}[i]=2i(i+1)+\frac{4}{3} i(i+1)(2i+1),\,\forall i \geq 0\,. }
\end{equation}
The number series yielding the number of NC interface bonds is  
\begin{equation}\label{eqn-07}
\mathbf{
N\mathrm{\mathbf{_{IF}^{octa}}}[i]= 4(i+1)^2,\,\forall i \geq 0\,. }
\end{equation}
The start cases of the number series ($i=0$) describe a single Si atoms with four interface bonds, presenting a silane molecule SiX$_4$, X = H, F, OH, NH$_2$, CH$_3$, etc. Octahedral \{111\}-faceted NCs described by these number series are shown in figure \ref{fig03} where atoms are color-coded in accord with their bond configuration.

\subsection{\label{dodeca}Dodecahedral \{110\}-Faceted Zinc-Blende Nanocrystals}
The interface termination and symmetry constraints of \{110\}-faceted dodecahedral NCs are significantly more complex as compared to \{001\}-faceted cubic NCs or \{111\}-faceted octahedral NCs. Figure \ref{fig05} shows that there are two different types of regular \{110\}-dodecahedral NCs which have the same form as the Brillouin zone of a body-centered cubic (bcc) lattice. These two dodecahedral NC classes will each have their own number series for $N\mathrm{_{NC}^{dod}}$, $N\mathrm{_{bnd}^{dod}}$ and $N\mathrm{_{IF}^{dod}}$. 
Due to their alternating occurence we name the class of NCs starting with the smallest NC as \emph{odd}, and the class of NCs starting with the second-smallest NC as \emph{even}. While it would be sufficient to use one NC class, we derive the number series for both classes which diminishes the size difference between adjacent NCs from 17.0 to 8.5 {\AA}. We will first describe the odd class of NCs in detail and then show solutions for the even class of NCs since latter findings use the same algorithm.

\begin{figure}[t!]
\begin{center}
\mbox{\scalebox{0.3764}{\includegraphics{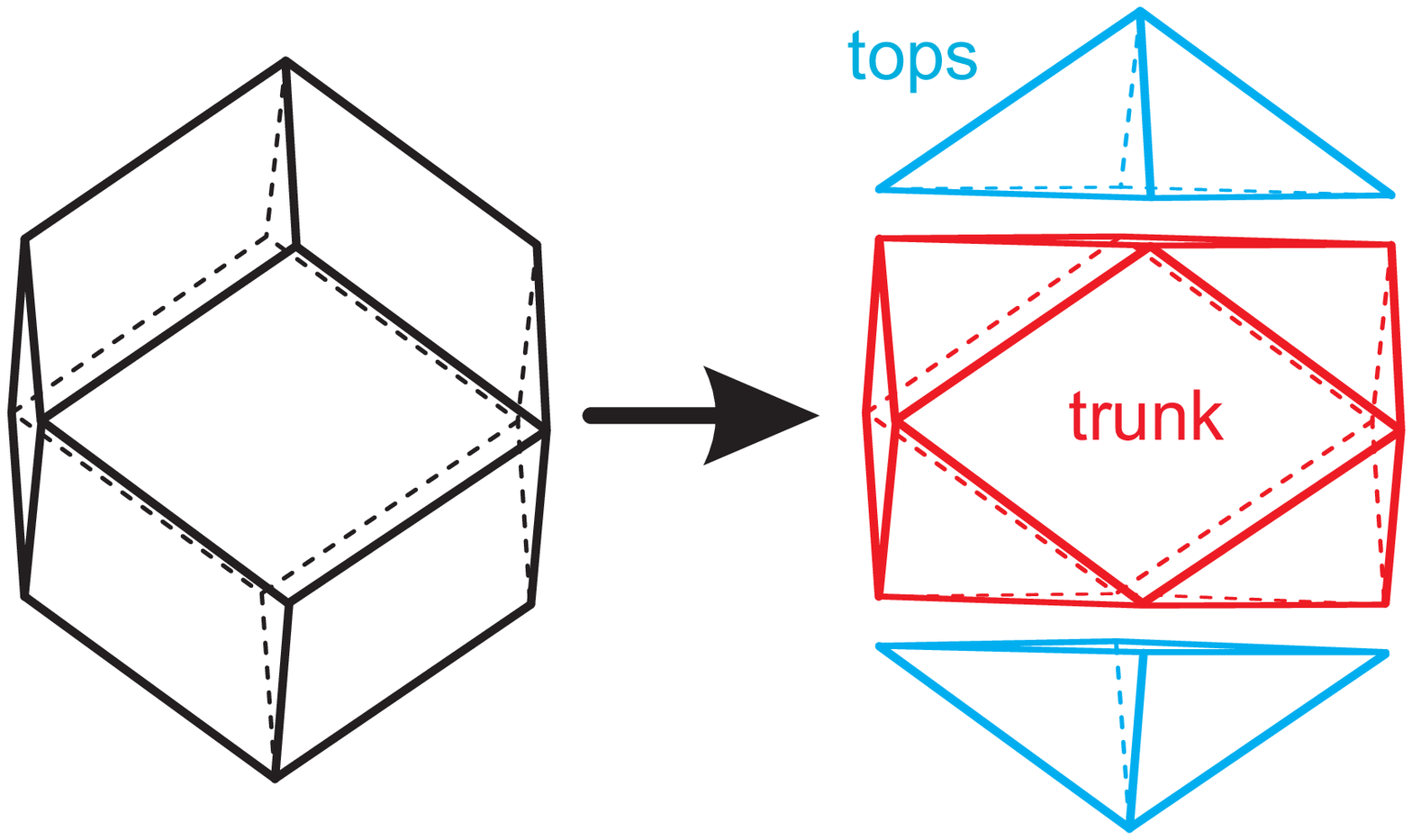}}}
\end{center}
\caption{{\bf Decomposition of \{110\}-faceted dodecahedra.} The trunk and top sections are required for deriving $N\mathrm{_{NC}}$ and $N\mathrm{_{bnd}}$ for each section type -- $N\mathrm{_{NC}^{dod,trnk}},\,N\mathrm{_{NC}^{dod,tops}}$ and $N\mathrm{_{bnd}^{dod,trnk}},\,N\mathrm{_{bnd}^{dod,tops}}$. The sum of the sections provides the total value of $N\mathrm{_{NC}}$ and $N\mathrm{_{bnd}}$, see equations \ref{eqn-10} and \ref{eqn-13} for the odd index number series.}
\label{fig04}
\end{figure}

\begin{figure*}[t!]
\begin{center}
\mbox{\scalebox{0.10568}{\includegraphics{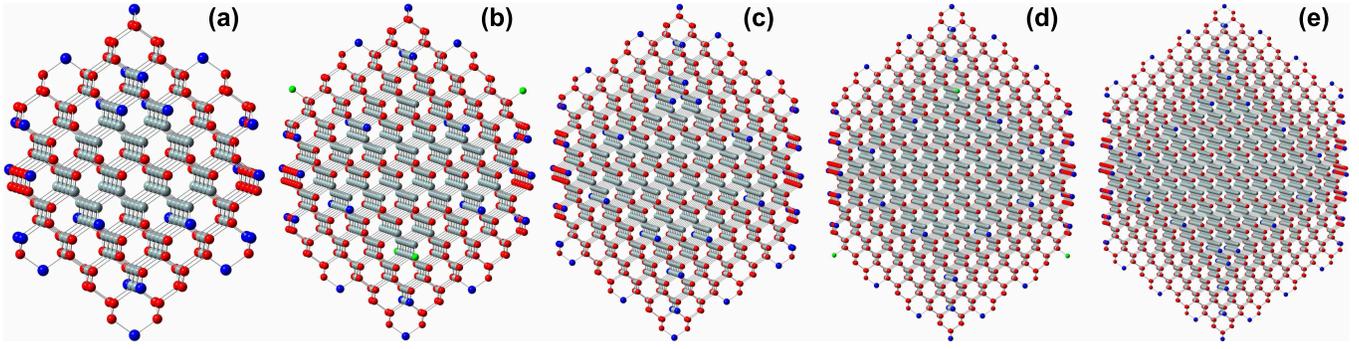}}}
\end{center}
\caption{{\bf Dodecahedral NCs with exclusive \{110\} facets described by equations \ref{eqn-16} and \ref{eqn-19} to \ref{eqn-23}.} Grey atoms have all bonds in NC. Red atoms have one interface bond, blue atoms have two interface bonds and green atoms have three interface bonds. Graphs (a) to (e) show NCs for $i=\mbox{1}_{\mathrm{odd}},\,\mbox{1}_{\mathrm{even}}$, $\mbox{2}_{\mathrm{odd}},\,\mbox{2}_{\mathrm{even}},$ and $\mbox{3}_{\mathrm{odd}}$, respectively. These NCs have to be split into two groups: NCs without atoms having three interface bonds (green) called odd due to first (a), third (c), fifth (e), etc. NC belonging into this group, and an even group (b, d) with four atoms having three interface bonds, \emph{cf.} table \ref{tab2}.}
\label{fig05}
\end{figure*}

The calculations of $N\mathrm{_{NC,odd}^{dod}}$ and $N\mathrm{_{bnd,odd}^{dod}}$ require a decomposition of dodecahedral NCs into a trunk and two tops, see figure \ref{fig04}. Each section type -- trunk and both tops -- can be described with third order differential schemes which can be transformed into $2^{\mathrm{nd}}$ order recursive number series with a linear term, scaling $\propto i^3$ as expected for a volume variable. Since the start values for $i=1$ and $i=2$ of both series are different (see equations \ref{eqn-08}, \ref{eqn-09} and \ref{eqn-11}, \ref{eqn-12}), they cannot be directly summed up into one equation for the respective  $N\mathrm{_{NC,odd}^{dod}}[i]$ or $N\mathrm{_{bnd,odd}^{dod}}[i]$. After we derived these number series in their recursive forms, we transform them into series which are an explicit function of $i$, allowing to merge the trunk and tops sections into one series $N\mathrm{_{NC,odd}^{dod}}$ and $N\mathrm{_{bnd,odd}^{dod}}$. 
\\
The 3$^{\mathrm{rd}}$ order differential scheme of NC atoms forming the trunk section is shown in table \ref{tab-01}. This scheme serves to derive equation \ref{eqn-08}, making use of the linear increase of the 2$^{\mathrm{nd}}$ differential quotient which is $d^2\,N\mathrm{_{NC,odd}^{dod,trnk}}[i]/d\,i^2 = 16(40[i+1]-27)$.
\begin{table*}[t!]
\caption{\label{tab-01} $3^{\mathrm{rd}}$ order differential scheme, see text for details.}\vspace*{0.2cm}
\renewcommand{\baselinestretch}{1.2}\small\normalsize
\begin{tabular}{rrrrr}
\rbT{$i$}\ &\ \rbT{$N\mathrm{_{NC,odd}^{dod,trnk}}[i]$}\ &\ $\frac{\displaystyle d\,N\mathrm{_{NC,odd}^{dod,trnk}}[i]}{\displaystyle d\,i}$\ &\ $\frac{\displaystyle d^2\,N\mathrm{_{NC,odd}^{dod,trnk}}[i]}{\displaystyle d\,i^2}$\ &\ 
    $\frac{\displaystyle d^3\,N\mathrm{_{NC,odd}^{dod,trnk}}[i]}{\displaystyle d\,i^3}$\ \\[0.1cm]  
  1 &246\hspace*{0.5cm}&&&\\ 
  2 &1336\hspace*{0.5cm}&\rbL{1090}\hspace*{0.7cm}&1488\hspace*{0.85cm}&\\ 
  3 &3914\hspace*{0.5cm}&\rbL{2578}\hspace*{0.7cm}&2128\hspace*{0.85cm}&
  \rbL{640}\hspace*{0.94cm}\\
  4 &8620\hspace*{0.5cm}&\rbL{4706}\hspace*{0.7cm}&2768\hspace*{0.85cm}&
  \rbL{640}\hspace*{0.94cm}\\
  5 &16094\hspace*{0.5cm}&\rbL{7474}\hspace*{0.7cm}&3408\hspace*{0.85cm}&
  \rbL{640}\hspace*{0.94cm}\\
  6 &26976\hspace*{0.5cm}&\rbL{10882}\hspace*{0.7cm}&4048\hspace*{0.85cm}&
  \rbL{640}\hspace*{0.94cm}\\
  7 &41906\hspace*{0.5cm}&\rbL{14930}\hspace*{0.7cm}&4688\hspace*{0.85cm}&
  \rbL{640}\hspace*{0.94cm}\\
  8 &61524\hspace*{0.5cm}&\rbL{19618}\hspace*{0.7cm}&$\ldots$\hspace*{0.85cm}&
  \rbL{$\ldots$}\hspace*{0.94cm}\\ 
  $\ldots$&$\ldots$\hspace*{0.5cm}&\rbL{$\ldots$}\hspace*{0.7cm}&&\rbL{$=$ const.}\hspace*{0.65cm}
  \end{tabular}
\renewcommand{\baselinestretch}{1.0}\small\normalsize
\end{table*}
Hence, the number of NC atoms forming the trunk of dodecahedral NCs is
\begin{eqnarray}\label{eqn-08}
  N\mathrm{_{NC,odd}^{dod,trnk}}[i]&=&
  16(40i-27)\,+\,2N\mathrm{_{NC,odd}^{dod,trnk}}[i-1]\\
  &&\,-\,N\mathrm{_{NC,odd}^{dod,trnk}}[i-2]\,,\forall i \geq 3\,;\nonumber\\
  &&\quad N\mathrm{_{NC,odd}^{dod,trnk}}[1] = 246,\ N\mathrm{_{NC,odd}^{dod,trnk}}[2] = 1336\,. \nonumber
\end{eqnarray}
Since $d^2\,N\mathrm{_{NC,odd}^{dod,trnk}}[i]/d\,i^2$ in the differential scheme relates to $N\mathrm{_{NC,odd}^{dod,trnk}}[i+1]$ (1488 to 3914 for $i=2$, etc.), the term $16(40[i+1]-27)$ translates to $16(40i-27)$ in equation \ref{eqn-08}.
We obtain the values for the two top sections along the same lines, yielding 
\begin{eqnarray}\label{eqn-09}
  N\mathrm{_{NC,odd}^{dod,tops}}[i]&=&
  16(8i-3)\,+\,2N\mathrm{_{NC,odd}^{dod,tops}}[i-1]\\
  &&\,-\,N\mathrm{_{NC,odd}^{dod,tops}}[i-2]\,,\forall i \geq 3\,;\nonumber\\
  &&\quad N\mathrm{_{NC,odd}^{dod,tops}}[1] = 92,\ N\mathrm{_{NC,odd}^{dod,tops}}[2] = 386\,.\nonumber
\end{eqnarray}
We add both series per $i$ to yield the total number of NC atoms: 
\begin{equation}\label{eqn-10}
  N\mathrm{_{NC,odd}^{dod}}[i]=N\mathrm{_{NC,odd}^{dod,trnk}}[i]+
  N\mathrm{_{NC,odd}^{dod,tops}}[i]\,,\ \forall i\geq 1\,.
\end{equation}

The same approach applies to the number of bonds connecting NC atoms. For the trunk section we get
\begin{eqnarray}\label{eqn-11}
  N\mathrm{_{bnd,\,odd}^{dod,trnk}}[i]&=&
  128(10i-7)\,+\,2N\mathrm{_{bnd,\,odd}^{dod,trnk}}[i-1]\\
  &&\,-\,N\mathrm{_{bnd,\,odd}^{dod,trnk}}[i-2]\,,\forall i \geq 3\,;\nonumber\\
  &&\quad N\mathrm{_{bnd,\,odd}^{dod,trnk}}[1] = 472,\ N\mathrm{_{bnd,\,odd}^{dod,trnk}}[2] = 2600\,.\nonumber
\end{eqnarray}
The series for the tops includes the bonds to the trunk, it is 
\begin{eqnarray}\label{eqn-12}
  N\mathrm{_{bnd,\,odd}^{dod,tops}}[i]&=&
  32(8i-5)\,+\,2N\mathrm{_{bnd,\,odd}^{dod,tops}}[i-1]\\
  &&\,-\,N\mathrm{_{bnd,\,odd}^{dod,tops}}[i-2]\,,\forall i \geq 3\,;\nonumber\\
  &&\quad N\mathrm{_{bnd,\,odd}^{dod,tops}}[1] = 112, N\mathrm{_{bnd,\,odd}^{dod,tops}}[2] = 572\,.\nonumber
\end{eqnarray}
Again, we add both sub-series to get the total number of bonds between NC atoms:
\begin{equation}\label{eqn-13}
  N\mathrm{_{bnd,odd}^{dod}}[i]=N\mathrm{_{bnd,odd}^{dod,trnk}}[i]+
  N\mathrm{_{bnd,odd}^{dod,tops}}[i]\,,\, \forall i\geq 1\,.
\end{equation}

We now convert the recursive forms of $N\mathrm{_{NC,odd}^{dod}}[i]$ and $N\mathrm{_{bnd,odd}^{dod}}[i]$ into number series depending explicitely on $i$. When unfolding $N\mathrm{_{NC,odd}^{dod,trnk}}[i]$ (equation \ref{eqn-08}) over $i$-terms, we obtain 
\begin{widetext}
\begin{eqnarray}\label{eqn-14}
&&N\mathrm{_{NC,odd}^{dod,trnk}}[i]=\\
&&16(40i-27)+2\Big[16(40[i-1]-27)+2\cdot 16(40[i-2]-27)-16(40[i-3]-27)\ldots\Big]-\nonumber\\
&&\Big[16(40[i-2]-27)+2\cdot 16(40[i-3]-27)+2\cdot 16(40[i-4]-27)-16(40[i-5]-27)\ldots\Big]\nonumber\\
&&=16\sum^{i}_{k=1}\Big[(i-k+1)(40k-27)\Big]\ \ +\ \ 2(17i+2)\nonumber\\
&&=16\left[(40i+13)\frac{i(i+1)}{2}-20\frac{i(i+1)(2i+1)}{3}\right]\ \ +\ \ 2(17i+2)\ ,\nonumber
\end{eqnarray}
\end{widetext}
whereby the last line used the terms $\sum\,k^2$ and $\sum\,k$ to replace the respective partial sum formula \cite{Zeid04}. The term $2(17i+2)$ is an offset derived from the intial values of equation \ref{eqn-08}. Since the start of the recursive series is fixed with $N\mathrm{_{NC,odd}^{dod,trnk}}[1]=246$ and $N\mathrm{_{NC,odd}^{dod,trnk}}[2]=1336$, we need to solve equation \ref{eqn-14} \emph{without the intial offset} for an intial value differential problem. We get $N\mathrm{_{NC,odd}^{dod,trnk}}=208, 1264, 3808, 8480\,\ldots$ for $i=1,\,2,\,3,\,4,\,\ldots$, respectively. Comparing these values to the ones obtained from equation \ref{eqn-08}, we get offsets of 38, 72, 106, $140,\,\ldots$ for $i=1,\,2,\,3,\,4,\,\ldots$, yielding $2(17i+2)$. In the same fashion, the explicit form of equation \ref{eqn-09} is
\begin{eqnarray}\label{eqn-15}
N\mathrm{_{NC,odd}^{dod,tops}}[i]&=&16\sum^{i}_{k=1}\Big[(i-k+1)(8k-3)\Big]\\
&&\ +\ 6(i+1)\nonumber\\
&=&16\left[(8i+5)\frac{i(i+1)}{2}-4\frac{i(i+1)(2i+1)}{3}\right]\nonumber\\
&&\ +\ 6(i+1)\ . \nonumber
\end{eqnarray}\\
Adding the series of equations \ref{eqn-14} and \ref{eqn-15} yields the \emph{odd} number series of atoms forming a dodecahedral \{110\}-terminated zb-NC:
\begin{eqnarray}\label{eqn-16}
\mathbf{N\mathrm{\mathbf{_{NC,odd}^{dod}}}[i]}&=&
\mathbf{16i\left[(i+1)(24i+9)-8(i+1)(2i+1)\right]}\nonumber\\
&&\mathbf{\ +\ 10(4i+1)\ .}
\end{eqnarray}
The number of bonds between atoms $N\mathrm{_{bnd,odd}^{dod}}[i]$ is derived in the same fashion, yielding 
\begin{eqnarray}\label{eqn-17}
N\mathrm{_{bnd,odd}^{dod,trnk}}[i]&=&128\sum^{i}_{k=1}\Big[(i-k+1)(10k-7)\Big]\\
&&\ +\ 8(10i+1)\nonumber\\
&=&128\left[(10i+3)\frac{i(i+1)}{2}-5\frac{i(i+1)(2i+1)}{3}\right]\nonumber\\
&&\ +\ 8(10i+1)\nonumber
\end{eqnarray}\\
for the trunk and 
\begin{eqnarray}\label{eqn-18}
N\mathrm{_{bnd,odd}^{dod,tops}}[i]&=&32\sum^{i}_{k=1}\Big[(i-k+1)(8k-5)\Big]\\
&&\ +\ 4(3i+1)\nonumber\\
&=&32\left[(8i+3)\frac{i(i+1)}{2}-4\frac{i(i+1)(2i+1)}{3}\right]\nonumber\\
&&\ +\ 4(3i+1)\nonumber
\end{eqnarray}\\
for the top sections. Adding both series, we get the explicit form
\begin{eqnarray}\label{eqn-19}
\mathbf{N\mathrm{_{bnd,odd}^{dod}}[i]}&=&\mathbf{16i\left[3(16i+5)(i+1)-16(i+1)(2i+1)\right]}\nonumber\\
&&\mathbf{\ +\ 4(23i+3)\ .} 
\end{eqnarray}\\
The number series for the bonds of the NCs connecting them to their environment can be solved in its explicit form: 
\begin{eqnarray}\label{eqn-20}
\mathbf{N\mathrm{\mathbf{_{IF,\,odd}^{dod}}}[i]}&=&
\mathbf{128i(i+1)-184i+112\,,}\nonumber\\
&&\mathbf{\ \forall i \geq 1\,. }
\end{eqnarray}

The even class of dodecahedral NCs follow derivations alike to the odd NC class, though coefficients for the number series and start elements are different. We only list the final explicit results:
\begin{eqnarray}\label{eqn-21}
\mathbf{N\mathrm{\mathbf{_{NC,even}^{dod}}}[i]}&=&
\mathbf{16i\left[(24i+21)(i+1)-8(i+1)(2i+1)\right]}\nonumber\\
&&\mathbf{\ +\ 88(i+1) }
\end{eqnarray}
\begin{widetext}
\begin{eqnarray}\label{eqn-22}
\mathbf{N\mathrm{\mathbf{_{bnd,even}^{dod}}}[i]}&=&
\mathbf{16i\left[3(16i+13)(i+1)-16(i+1)(2i+1)\right]}\nonumber\\
&&\mathbf{\ +\ 140(i+1)\ . }
\end{eqnarray}
\end{widetext}
Again, the number series of NC interface bonds is explicit:
\begin{equation}\label{eqn-23}
\mathbf{N\mathrm{\mathbf{_{IF,\,even}^{dod}}}[i]=128i(i+1)-56i+136\,,\ 
\forall i \geq 1\ . }
\end{equation}
Minor oscillations of $N\mathrm{_{IF}^{dod}}[i]$ occur for small NCs due to alternating odd and even terms, \emph{cf.} Figs. \ref{fig04}b and \ref{fig11}.  These originate from different interface bond configurations for odd and even NC series, see figure \ref{fig05}. The origin of this irregular behavior is found in the less perfect termination of \{110\} facets due to the more complex geometry of \{110\} surfaces in compound with their inter-facet angles being obtuse at all edges. These imperfections are less dominant for small dodecahedra due to the small area of rhomboid \{110\}-facets since there are less chains of \{110\}-oriented surface atoms. These chains have an atom with two interface bonds at each end which contributes to $N\mathrm{_{IF}^{dod}}[i]$. 
\begin{table}[t!]
\caption{First members of geometrical series of the number of atoms contained in a NC $N_{\mathrm{NC}}$, the number of bonds between such atoms $N\mathrm{_{bnd}}$, the interface bonds of the NC $N\mathrm{_{IF}}$ for zb-NCs with exclusive \{001\} faceeting (cubes), \{111\}-faceting (octahedra) and \{110\}-faceting (dodecahedra), and the equivalent diameter $d_{\mathrm{NC}}$ when its volume is considered as sphere to allow for comparison between different NC shapes. An atomic volume for Si of $V_{\mathrm{atom}}=20.024$ {\AA}$^3$ was used.} \label{tab2}\vspace*{0.1cm} 
\renewcommand{\baselinestretch}{1.5}\small
  \begin{tabular}{|l||c|c|c|c|c|c|c|c}
    \hline
    $i$&1&2&3&4&5&6&7\\ \hline\hline
  \multicolumn{8}{c}{cubic shape, \{001\} surfaces}\\ \hline
  $N_{\mathrm{NC}}$&8&64&216&512&1000&1728&2744\\ \hline
  $N\mathrm{_{bnd}}$&7&86&333&844&1715&3042&4921\\ \hline
  $N\mathrm{_{IF}}$&18&84&198&360&570&828&1134\\ \hline
  $d_{\mathrm{NC}}$ [{\AA}]&6.74&13.5&20.2&27.0&33.7&40.4&47.2\\ \hline Figure&\ref{fig02}a&\ref{fig02}b&\ref{fig02}c&\ref{fig02}d&\ref{fig02}e&\ref{fig02}f&\\ \hline\hline
  \multicolumn{8}{c}{octahedral shape, \{111\} surfaces}\\ \hline
  $N_{\mathrm{NC}}$&10&35&84&165&286&455&680\\ \hline
  $N\mathrm{_{bnd}}$&12&52&136&280&500&812&1232\\ \hline
  $N\mathrm{_{IF}}$&16&36&64&100&144&196&256\\ \hline
  $d_{\mathrm{NC}}$ [{\AA}]&7.26&11.0&14.8&18.5&22.2&25.9&29.6\\ \hline  Figure&\ref{fig03}a&\ref{fig03}b&\ref{fig03}c&\ref{fig03}d&\ref{fig03}e&\ref{fig03}f&\ref{fig03}g\\ \hline\hline
  \multicolumn{8}{c}{dodecahedral shape, \{110\} surfaces}\\ \hline
   \hline
    $i$&$1_{\mathrm{odd}}$&$1_{\mathrm{even}}$&$2_{\mathrm{odd}}$
    &$2_{\mathrm{even}}$&$3_{\mathrm{odd}}$&$3_{\mathrm{even}}$
    &$4_{\mathrm{odd}}$\\ \hline\hline
  $N_{\mathrm{NC}}$&338&848&1722&3048&4930&7456&10730\\ \hline
  $N\mathrm{_{bnd}}$&584&1528&3172&5700&9312&14192&20540\\ \hline
  $N\mathrm{_{IF}}$&184&336&512&792&1096&1504&1936\\ \hline
  $d_{\mathrm{NC}}$ [{\AA}]&23.5&31.9&40.4&48.9&57.4&65.8&74.3\\ \hline
Figure&\ref{fig05}a&\ref{fig05}b&\ref{fig05}c&\ref{fig05}d&\ref{fig05}e&&\\ \hline    
  \end{tabular}
\renewcommand{\baselinestretch}{1.0}\small\normalsize
\end{table}

\subsection{Tetrahedral Zinc-Blende Nanocrystals}\label{Tetra}
With exclusive \{111\} faceting of zb-NCs, the resulting tetrahedral NCs can also be seen as regular pyramids with three sides. For the number of NC atoms forming the tetrahedral \{111\} zb-NC we get 
\begin{eqnarray}\label{eqn-24}
\mathbf{N\mathrm{\mathbf{_{NC}^{tetra}}}[i]}&=&
\mathbf{\frac{1}{6}i(i+1)(2i+1)+(i+1)^2\,,}\nonumber\\ 
&&\mathbf{\ \forall i\geq 0\,, }
\end{eqnarray}
The bonds between such NC atoms are described by 
\begin{eqnarray}\label{eqn-25}
\mathbf{N\mathrm{\mathbf{_{bnd}^{tetra}}}[i]}&=&
\mathbf{\frac{1}{3}i(i+1)(2i+1)+i(i+1)\,,}\nonumber\\
&&\mathbf{\ \forall i \geq 0\,. }
\end{eqnarray}
The number of interface bonds connecting these NCs to their embedding matrix or attached ligands is given by 
\begin{equation}\label{eqn-26}
\mathbf{N\mathrm{\mathbf{_{IF}^{tetra}}}[i]=2(i+1)(i+2)\,,\ \forall i 
\geq 0\,. }
\end{equation}\unboldmath

Equation \ref{eqn-26} deviates from equations \ref{eqn-07} and \ref{eqn-29} because base and side areas of a tetragonal zb-NC with \{111\} faceting are identical. 
\begin{figure*}[t!]
\begin{center}
\mbox{\scalebox{0.0542}{\includegraphics{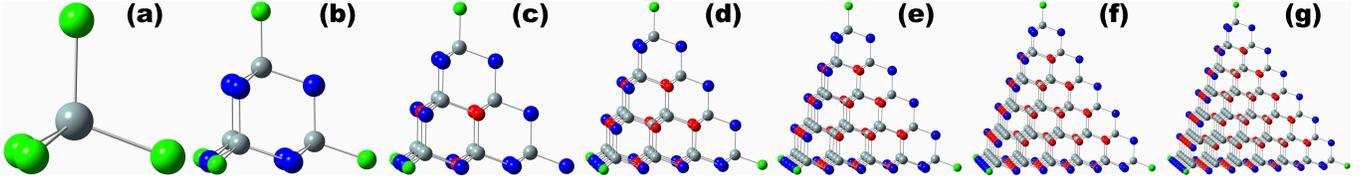}}}
\end{center}
\caption{{\bf Tetrahedral NCs with exclusive \{111\} faceting as described by equations \ref{eqn-24} to \ref{eqn-26}.} For atoms colors, see figure \ref{fig05}. Graphs (a) to (g) show tetrahedral NCs for $i=\mbox{1,\,2,\,3,\,4,\,5,\,6}$ and 7, respectively, see also table \ref{tab3}.}
\label{fig06}
\end{figure*}
\begin{figure*}[t!]
\begin{center}
\mbox{\scalebox{0.08769}{\includegraphics{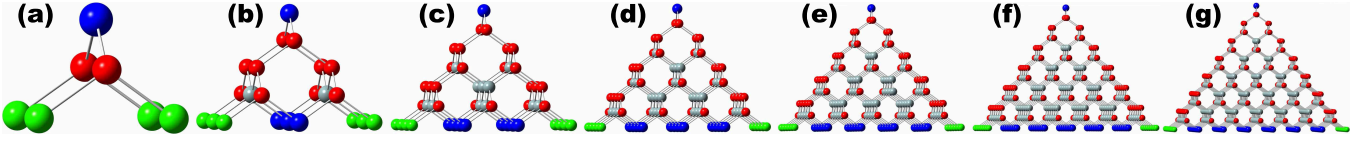}}}
\end{center}
\caption{{\bf Pyramidal NCs with \{001\} base surface and \{111\} side surfaces as described by equations \ref{eqn-27} to \ref{eqn-29}.} Graphs (a) to (g) show pyramid-shape NCs for $i=\mbox{1,\,2,\,3,\,4,\,5,\,6}$ and 7, respectively, see also table \ref{tab3}.}
\label{fig07}
\end{figure*}

\subsection{\label{Pyra}\{111\}-\{001\} Pyramidal Zinc-Blende Nanocrystals}
Pyramidal NCs consist of four triangular \{111\} side facets and a square \{001\}-oriented base. Such NCs are relevant in particular for epitaxial growth of III-V quantum dots (QDs) \cite{Grun95,Bai04}. The number of atoms forming the NC are 
\begin{eqnarray}\label{eqn-27}
\mathbf{N\mathrm{\mathbf{_{NC}^{pyra}}}[i]}&=&
\mathbf{\frac{1}{6}i(i+1)\big[2(2i+1)+9\big]+i+1\,,}\nonumber\\
&&\mathbf{\ \forall i\geq 0\,. }
\end{eqnarray} 
The bonds between such NC atoms are described by 
\begin{equation}\label{eqn-28}
\mathbf{N\mathrm{\mathbf{_{bnd}^{pyra}}}[i]=\frac{2}{3}i(i+1)(2i+1)+i(i+1)\,,\ \forall i \geq 0\,. }
\end{equation}
The number of NC interface bonds is given by 
\begin{equation}\label{eqn-29}
\mathbf{
 N\mathrm{\mathbf{_{IF}^{pyra}}}[i]=4(i+1)^2,\,\forall i \geq 0\,, }
\end{equation}
The identity of equations \ref{eqn-07} and \ref{eqn-29} shows the resemblance of symmetry arguments of both series which is underlined by the start term ($i=0$) describing SiX$_4$ in  both cases.
\begin{table}[t!]
\caption{First members of geometrical series for tetrahedral zb-NCs with exclusive \{111\}-faceting and pyramidal zb-NCs with \{111\}-oriented side areas and \{001\} base, see table \ref{tab2} for further details.}\label{tab3}
\vspace*{0.1cm}
\renewcommand{\baselinestretch}{1.5}\small
  \begin{tabular}{|l||c|c|c|c|c|c|c|c}\hline
  $i$&1&2&3&4&5&6&7\\ \hline\hline
  \multicolumn{8}{c}{tetragonal shape, \{111\} surfaces}\\ \hline
  $N_{\mathrm{NC}}$&5&14&30&55&91&140&204\\ \hline
  $N\mathrm{_{bnd}}$&4&16&40&80&140&224&336\\ \hline
  $N\mathrm{_{IF}}$&12&24&40&60&84&112&144\\ \hline
  $d_{\mathrm{NC}}$ [{\AA}]&5.8&8.1&10.5&12.8&15.2&17.5&19.8\\ \hline
Figure&\ref{fig06}a&\ref{fig06}b&\ref{fig06}c&\ref{fig06}d&\ref{fig06}e&\ref{fig06}f&\ref{fig06}g\\ \hline\hline
   \multicolumn{8}{c}{pyramidal shape, \{001\} base, \{111\} sides}\\ \hline
  $N_{\mathrm{NC}}$&7&22&50&95&161&252&372\\ \hline
  $N\mathrm{_{bnd}}$&6&26&68&140&250&406&616\\ \hline
  $N\mathrm{_{IF}}$&16&36&64&100&144&196&256\\ \hline
  $d_{\mathrm{NC}}$ [{\AA}]&6.4&9.4&12.4&15.4&18.3&21.3&24.2\\ \hline
Figure&\ref{fig07}a&\ref{fig07}b&\ref{fig07}c&\ref{fig07}d&\ref{fig07}e&\ref{fig07}f&\ref{fig07}g\\ \hline   
  \end{tabular}
\renewcommand{\baselinestretch}{1.0}\small\normalsize
\end{table}

\subsection{\label{quatro111}Quatrodecahedral Zinc-Blende Nanocrystals with Dominant\\ \{111\}-Faceting (Truncated Octahedra)}
\begin{figure}[h!]
\begin{center}
\mbox{\scalebox{0.1225}{\includegraphics{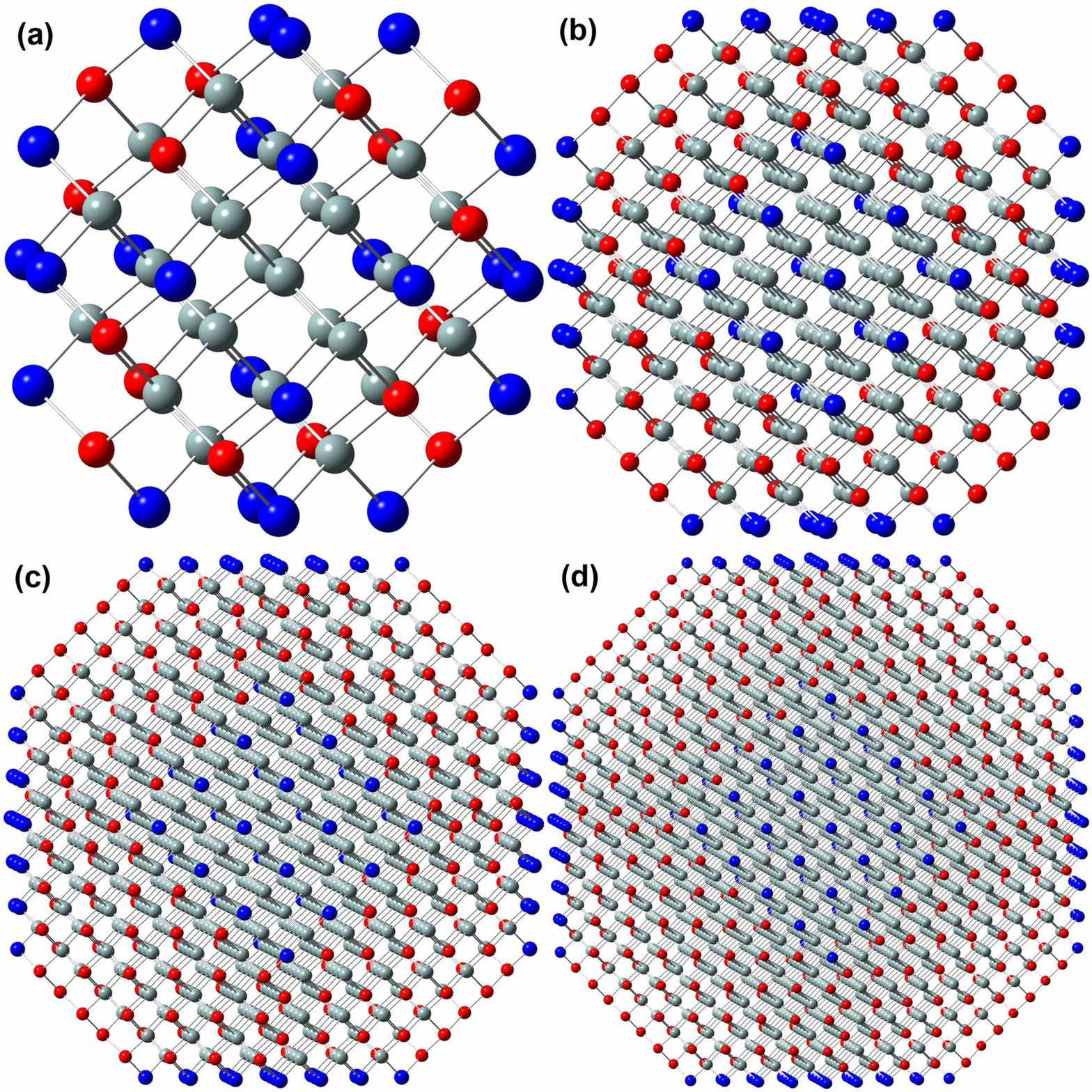}}}\vspace*{0.2cm}\\
\mbox{\scalebox{0.1226}{\includegraphics{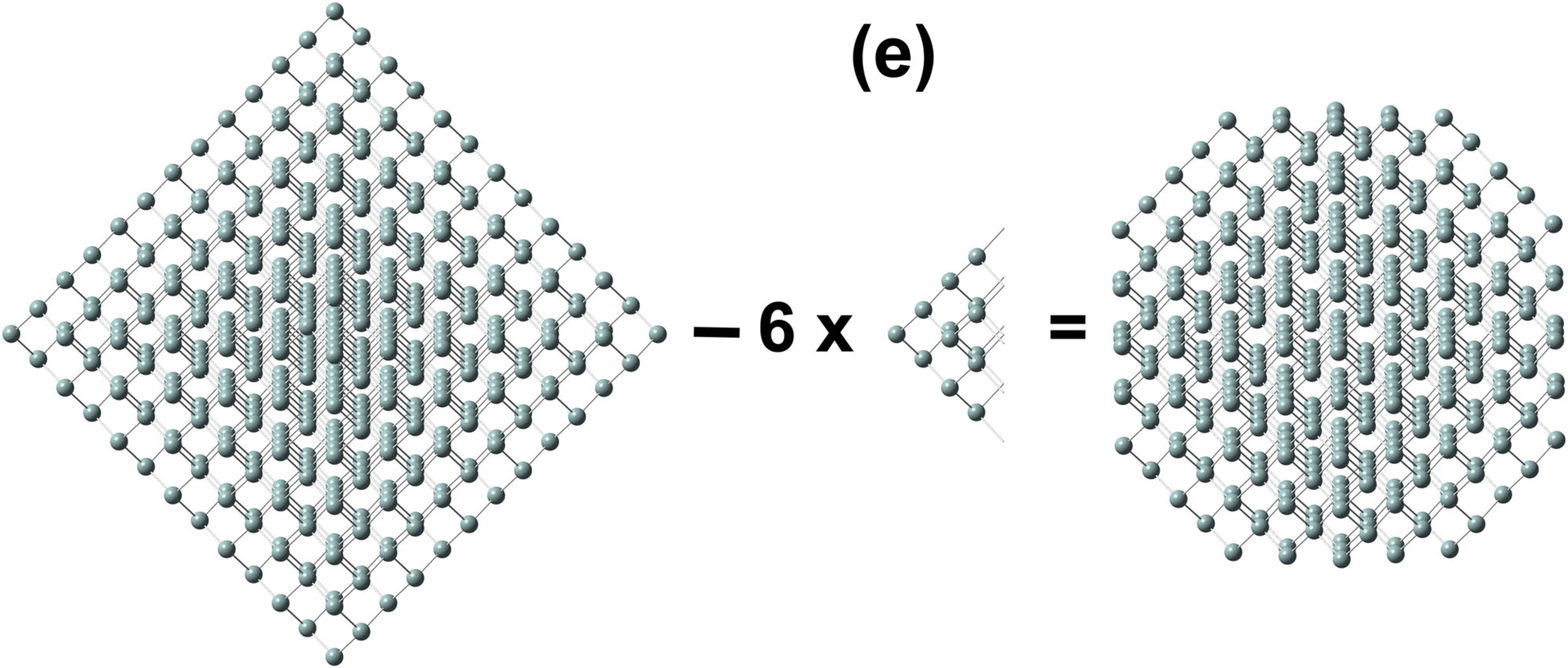}}}
\end{center}
\caption{{\bf Quatrodecahedral Zinc-Blende Nanocrystals with Dominant \{111\}-faceting as described by equations \ref{eqn-30} to \ref{eqn-32} (a, b, c, d).} For atoms colors, see figure \ref{fig05}. Graphs (a) to (d) show NCs for $i=\mbox{1,\,2,\,3}$ and 4, respectively, see table \ref{tab4}. {\bf Decomposition of \{111\}-faceted octahedron into six pyramidal \{001\}-\{111\} NCs and a dominantly \{111\}-faceted quatrodecahedral NC (e).} The bottom atom layers of the pyramidal NCs are part of the \{111\}-dominated quatrodecahedral NC to achieve its symmetry and shape.}
\label{fig08}
\end{figure}
\begin{figure}[t!]
\begin{center}
\mbox{\scalebox{0.1136}{\includegraphics{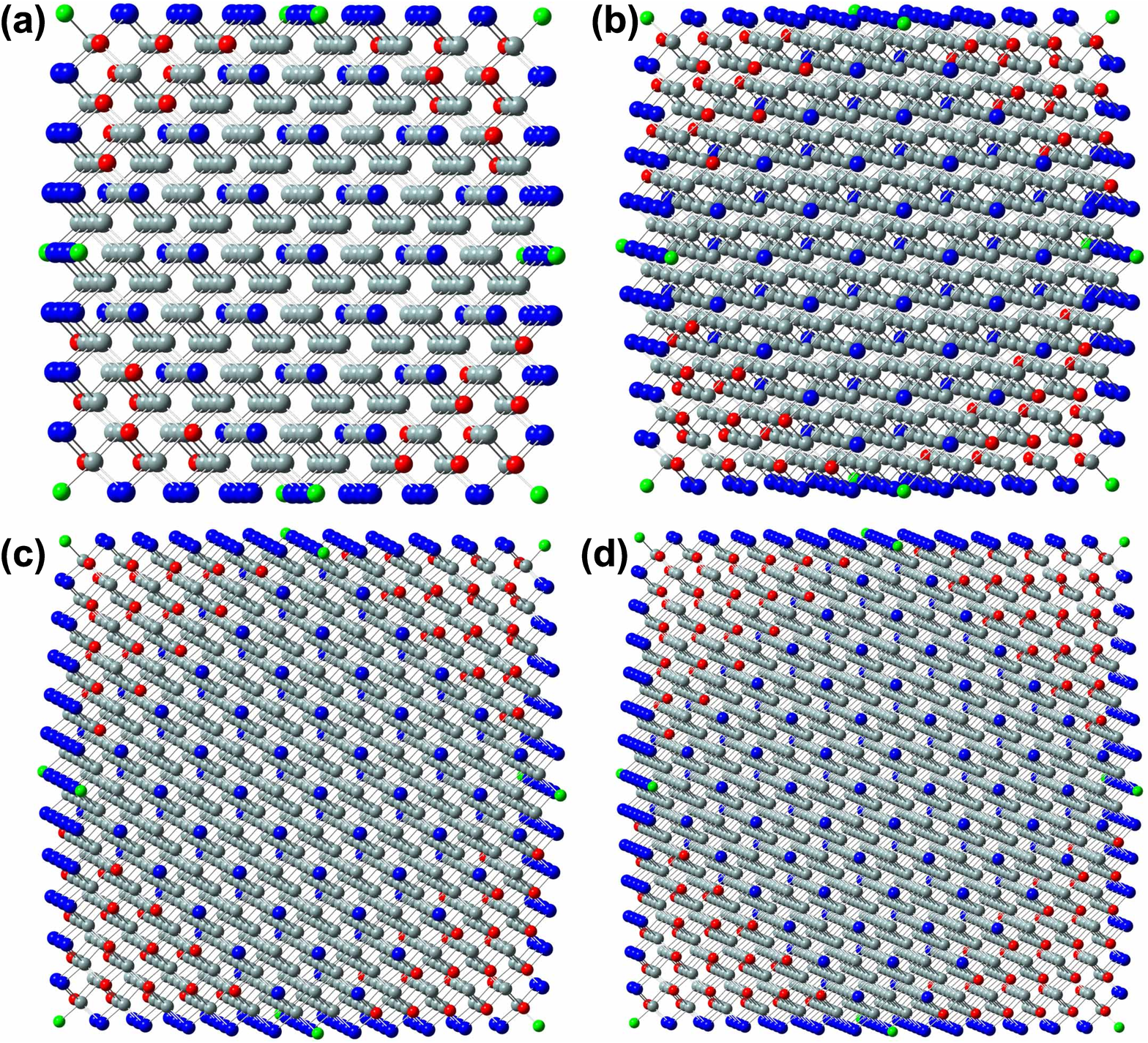}}}
\end{center}
\caption{{\bf Quatrodecahedral Zinc-Blende Nanocrystals with dominant \{001\}-faceting as described by equations \ref{eqn-35} to \ref{eqn-37}.} For atoms colors, see figure \ref{fig05}. Graphs (a) to (d) show NCs for $i=\mbox{1,\,2,\,3}$ and 4, respectively, see table \ref{tab4}.}
\label{fig09}
\end{figure}
The \{111\}-dominated quatrodecahedra are NCs which are limited by eight regular \{111\}-faceted hexagons and six \{001\}-faceted squares, see figure \ref{fig08}a to d. Quantities for these NCs can be derived from truncating \{111\}-faceted octahedra (Section \ref{Octa}) at their six corners, making use of the number series we obtained for the pyramids (Section \ref{Pyra}), \emph{cf.} figure \ref{fig08}e. We start again with the series yielding the number of NC atoms: 
\begin{eqnarray}\label{eqn-30}
\mathbf{N\mathrm{\mathbf{_{NC}^{q.111}}}[i]}&=&
\mathbf{9i(2i+1)^2+(2i+1)}\\
&&\mathbf{\,-\,i(4i+5)(i+1)\,,\ \forall i \geq 1\,. }\nonumber
\end{eqnarray}
The bonds between such NC atoms are described by 
\begin{eqnarray}\label{eqn-31}  \mathbf{N\mathrm{\mathbf{_{bnd}^{q.111}}}[i]}&=&\mathbf{2i(3i+1)(12i+5)}\\
&&\mathbf{\,-\big[ 4i(i+1)(2i+1)+6i(i+1)\big]\,,\ \forall i \geq 1\,.}\nonumber
\end{eqnarray}
The number of interface bonds which connect these NCs to an embedding matrix or attached ligands is given by 
\begin{eqnarray}\label{eqn-32}
\mathbf{
 N\mathrm{\mathbf{_{IF}^{q.111}}}[i]}&=&\mathbf{(6i+2)^2\,,\ \forall i \geq 1\,. }
\end{eqnarray}\\
For facilitating interpretations of electron paramagnetic resonance (EPR) data of SiO$_2$-embedded Si NCs \cite{Stes08} as discussed in section \ref{Applicat}, we decompose $N\mathrm{_{bnd}^{q.111}}[i]$ into a fraction of bonds originating from the six \{001\} facets where each atom has two interface bonds (blue atoms in figure \ref{fig08}a to d)
\begin{eqnarray}\label{eqn-33}
 N\mathrm{_{IF,\langle 001\rangle}^{q.111}}[i]&=&12(i+1)^2\,,\ \forall i \geq 1\,,
\end{eqnarray}\\
and another fraction originating from \{111\} facets where each atom has one interface bond (red atoms in figure \ref{fig08}a to d)
\begin{eqnarray}\label{eqn-34}
 N\mathrm{_{IF,\langle 111\rangle}^{q.111}}[i]&=&8(3i^2 -1)\,,\ \forall i \geq 1\,.
\end{eqnarray} 

\subsection{\label{quatro001}Quatrodecahedral Zinc-Blende Nanocrystals with Dominant\\ \{001\}-Faceting (Truncated Cubes)}
Here, the situation is more complex. The truncated corners of the cubic NC have a low symmetry as evident from different triangular facets having \{001\} orientations for side areas with angles of 45, 45 and 90$^{\circ}$ and a \{111\}-oriented base with three 60$^{\circ}$ angles. These truncated corners 
cannot be decribed by pyramids (Section \ref{Pyra}) or tetrahedra (Section \ref{Tetra}). As illustrated in section \ref{dodeca} for dodecahedral \{110\}-terminated zb-NCs, these NCs can be described with third order differential schemes which can be transformed into $2^{\mathrm{nd}}$ order recursive number series and eventually into an explicit form of $N\mathrm{_{NC}^{q.001}}[i]$ and $N\mathrm{_{bnd}^{q.001}}[i]$:
\begin{eqnarray}\label{eqn-35}
\mathbf{N\mathrm{\mathbf{_{NC}^{q.001}}}[i]}&=&
\mathbf{(2i+7)\Bigg[(i+4)^2+\frac{2}{3}(i+3)\left(i+\frac{5}{2}\right)\Bigg]}\nonumber\\[0.2cm]
&&\mathbf{\,+\,2(i+3)\left[(i+4)^2+i\right]}\nonumber\\
&&\mathbf{\,+\,\frac{4}{3}(i+1)(i+2)(i+3)\,,\ \forall i \geq 1}
\end{eqnarray}\\
and
\begin{widetext}
\begin{eqnarray}\label{eqn-36}
\mathbf{N\mathrm{\mathbf{_{bnd}^{q.001}}}[i]}&=&\\
&&\mathbf{4\Bigg[6(i+3)^2+(i+1)\left(i\left\{\frac{10}{3}i+\frac{11}{3}\right\}+
  \frac{9}{2}i+5\right)+7(2i+3)(i+2)\Bigg]\,,\ \forall i \geq 1 .\nonumber}
\end{eqnarray}
\end{widetext}
The number of interface bonds for these NCs is given by 
\begin{eqnarray}\label{eqn-37}
\mathbf{N\mathrm{\mathbf{_{IF}^{q.001}}}[i]}&=&\mathbf{4(2i+7)^2\,,\ \forall i \geq 1\,.}
\end{eqnarray}\\
As for dominantly \{001\}-faceted quatrodecahedra, we decompose $N\mathrm{_{IF}^{q.001}}[i]$ into \{001\} and \{111\} partitions to facilitate data interpretation for EPR measurements of SiO$_2$-embedded Si NCs \cite{Stes08}, see section \ref{Applicat}.
The \{001\} fraction of bonds originates from the six \{001\} facets where blue atoms have two interface bonds, \emph{cf.} figure \ref{fig09}, and 12 green atoms with three interface bonds, the latter accounting for the 36 added in equation \ref{eqn-38}. 
\begin{eqnarray}\label{eqn-38}
 N\mathrm{_{IF,\langle 001\rangle}^{q.001}}[i]&=&12\big[(i+4)^2 -4\big]+36 \nonumber\\
 &=&12\big[(i+4)^2 -1\big]\,,\ \forall i \geq 1\,.
\end{eqnarray}\\
The fraction which originates from \{111\} facets where each atom has one interface bond (red atoms in figure \ref{fig09}) is described by
\begin{eqnarray}\label{eqn-39}
 N\mathrm{_{IF,\langle 111\rangle}^{q.001}}[i]&=&4(i+2)^2 ,\,\forall i \geq 1\,.
\end{eqnarray} 
\begin{table}[t!]
\caption{First members of geometrical series for quatrodecahedral zb-NCs with dominant \{111\}- and \{001\}-faceting. Partitions $N\mathrm{_{IF,\langle 001\rangle}}[i]$ and $N\mathrm{_{IF,\langle 111\rangle}}[i]$ are discussed in section \ref{Applicat}. For further details see table \ref{tab2}.}
\label{tab4}\vspace*{0.1cm}
\renewcommand{\baselinestretch}{1.5}\small
  \begin{tabular}{|l||c|c|c|c|c|c|c|c}
    \hline
    $i$&1&2&3&4&5&6&7\\ \hline\hline
  \multicolumn{8}{c}{quatrodecahedral shape, \{111\}-dominant}\\ \hline
  $N_{\mathrm{NC}}$&66&377&1126&2505&4706&7921&12342\\ \hline
  $N\mathrm{_{bnd}}$&100&656&2052&4672&8900&15120&23716\\ \hline
  $N\mathrm{_{IF}}$&64&196&400&676&1024&1444&1936\\ \hline
  $N\mathrm{_{IF,\langle 001\rangle}}$&48&108&192&300&432&588&768\\ \hline
  $N\mathrm{_{IF,\langle 111\rangle}}$&12&88&208&376&592&856&1168\\ \hline
  $d_{\mathrm{NC}}$ [{\AA}]&13.6&24.3&$\!35.1\!$&$\!45.8\!$&$\!56.5\!$&67.2&77.9\\ \hline
    Figure&\ref{fig08}a&\ref{fig08}b&\ref{fig08}c&\ref{fig08}d&&&\\ \hline\hline
  \multicolumn{8}{c}{quatrodecahedral shape, \{001\}-dominant}\\ \hline
  $N_{\mathrm{NC}}$&549&1021&1707&2647&3881&5449&7391\\ \hline
  $N\mathrm{_{bnd}}$&936&1800&3076&4844&7184&10176&13900\\ \hline
  $N\mathrm{_{IF}}$&324&484&676&900&1156&1444&1764\\ \hline
  $N\mathrm{_{IF,\langle 001\rangle}}$&288&420&576&756&960&1188&1440\\ \hline
  $N\mathrm{_{IF,\langle 111\rangle}}$&36&64&100&144&196&256&324\\ \hline
  $d_{\mathrm{NC}}$ [{\AA}]&27.6&33.9&40.3&46.6&53.0&59.3&65.7\\ \hline
Figure&\ref{fig09}a&\ref{fig09}b&\ref{fig09}c&\ref{fig09}d&&&\\ \hline  
  \end{tabular}
\renewcommand{\baselinestretch}{1.0}\small\normalsize
\end{table}

\section{\label{results}Results}
\begin{figure}[t!]
\begin{center}
\mbox{\scalebox{0.3002}{\includegraphics{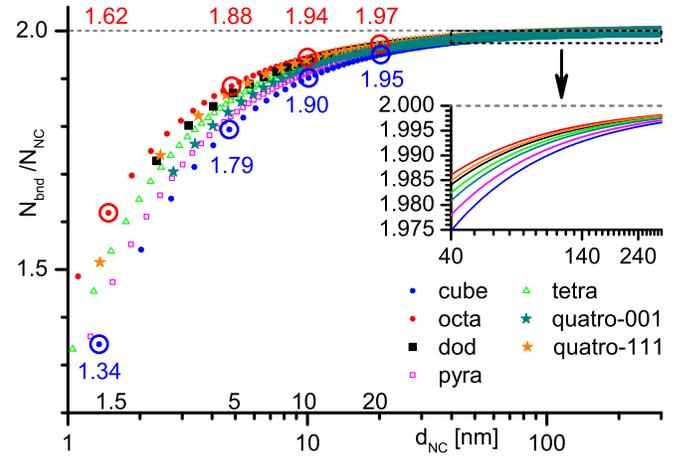}}}
\end{center}
\caption{{\bf Ratio of bonds per NC atom within NC $\mathbf{N\mathrm{\mathbf{ _{bnd}}}/N\mathrm{\mathbf{_{NC}}}}$}, shown for cubic (cube), octahedral (octa), dodecahedral (dod), pyramidal (pyra), tetrahedral (tetra), \{001\}-dominated (quatro-001) and \{111\}-dominated (quatro-111) quatrodecahedral NCs. For size values at bottom and associated $N\mathrm{_{bnd}}/N\mathrm{_{NC}}$ values as function of NC shape, see Section \ref{Applicat} and figure 7 in \cite{Steg09}. Inset zooms up range for 40 to 300 nm NC size to show asymptotic behavior. Lines were used to minimize obstruction.}
\label{fig10}
\end{figure}

The ratio $N\mathrm{_{bnd}}(d_{\mathrm{NC}}[i])/N_{\mathrm{NC}}(d_{\mathrm{NC}}[i])$ provides the number of bonds per atom within the zb-NC, the asymptotic limit $\lim_{i\to\infty}(N\mathrm{_{bnd}[i]}/N_{\mathrm{NC}}[i])=2$ describes the bulk case: The considered atom has four bonds, each shared with a first next neighbour (1-nn) atom. A finite crystal has outermost bonds missing as these connect the NC with its environment which results in $N\mathrm{_{bnd}}/N_{\mathrm{NC}}$ increasingly declining below 2 with shrinking $d_{\mathrm{NC}}[i]$, \emph{cf.} figure \ref{fig10}.

\begin{figure}[t!]
\begin{center}
\mbox{\scalebox{0.3002}{\includegraphics{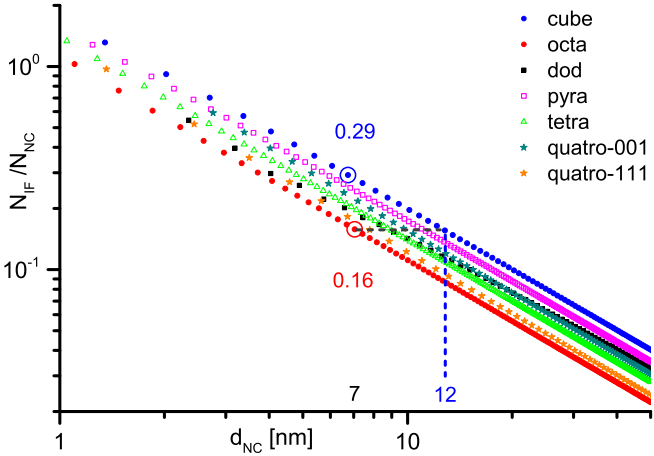}}}
\end{center}
\caption{{\bf Ratio of interface bonds per NC to number of NC atoms  $\mathbf{N\mathrm{\mathbf{_{IF}}}/N\mathrm{\mathbf{_{NC}}}}$.} The values of $N\mathrm{_{IF}}/N\mathrm{_{NC}}$ shown for cubic and octahedral NCs at 7 nm refer to the minimum size limit below which a strong dielectric embedding or coating the NC will dominate its electronic properties \cite{Koe14}. Cubic high symmetry zb-NCs are dominated by the dielectric up to $d_{\mathrm{NC}}=12$ nm (dashed blue line), lower symmetry NCs like Si cubicle fins of fin-FETs are dominated up to even bigger $d_{\mathrm{NC}}$ due to cubicle shape driving up $N\mathrm{_{IF}}/N\mathrm{_{NC}}$. For legend description see figure \ref{fig10}.}
\label{fig11}
\end{figure} 
Depending on the NC form and faceting,  $N\mathrm{_{bnd}}(d_{\mathrm{NC}}[i])/N_{\mathrm{NC}}(d_{\mathrm{NC}}[i])$ can vary in particular for small NCs. 

Exclusively \{111\}-faceted octahedra have the highest value per NC size $d_{\mathrm{NC}}$ by combining a low volume-to-surface ratio with a minimum surface bond density, \emph{cf.} table \ref{tab-00}. Quatrodecahedra with dominant \{111\}-faceting and square \{001\} facets have the next highest ratio $N\mathrm{_{bnd}}(d_{\mathrm{NC}}[i])/N_{\mathrm{NC}}(d_{\mathrm{NC}}[i])$. The latter facets cause an increase of $N\mathrm{_{IF}}[i]$. Both NCs types are closely followed by dodecahedral NCs with exclusive \{110\}-faceting. Although dodecahedra have a higher volume-to-surface ratio as compared to octahedra, 
their more complex \{110\} faceting results in a less perfect surface termination. Thereby, the number of bonds within the NC is decreased, the missing bonds are added to the number of interface bonds $N\mathrm{_{IF}}[i]$. Dodecahedra are followed by \{111\}-tetrahedra which have less favourable (smaller) volume-to-surface ratios. However, the minimum bond density of \{111\} surfaces (see table \ref{tab-00}) yields high $N\mathrm{_{bnd}}(d_{\mathrm{NC}}[i])/N_{\mathrm{NC}}(d_{\mathrm{NC}}[i])$ values. These NCs are followed by pyramids with \{001\} base and \{111\} side facets. We obtain the lowest values for $N\mathrm{_{bnd}}(d_{\mathrm{NC}}[i])/N_{\mathrm{NC}}(d_{\mathrm{NC}}[i])$ for \{001\}-faceted cubic NCs. This is straightforward to show by their unfavourable volume-to-surface ratio combined with the maximum bond density of \{001\}-facets, \emph{cf.} table \ref{tab-00}.

The ratio $N\mathrm{_{IF}}(d_{\mathrm{NC}}[i])/N_{\mathrm{NC}}(d_{\mathrm{NC}}[i])$ yields the number of interface bonds per NC atom, see figure \ref{fig11}. This key parameter quantifies electronic phenomena occurring across NC interfaces, see section \ref{Applicat}. It follows the opposite trend as discussed for $N\mathrm{_{bnd}}(d_{\mathrm{NC}}[i])/N_{\mathrm{NC}}(d_{\mathrm{NC}}[i])$: Any bond not available for connecting NC atoms occurs at an interface. The irregular behavior of \{110\}-faceted dodecahedra also occurs for $N\mathrm{_{IF}}(d_{\mathrm{NC}}[i])/N_{\mathrm{NC}}(d_{\mathrm{NC}}[i])$ for reasons discussed above. Elongated zb-NCs have a higher ratio $N\mathrm{_{IF}}/N\mathrm{_{NC}}$ which results in bigger $d_{\mathrm{NC}}$ values up to which the embedding dielectric dominates electronic and optical NC properties. This finding is crucial for \{001\}-terminated Si cubicles as encountered in fin-FETs to exploit nanoscopic phenomena such as ultrathin SiO$_2$ and Si$_3$N$_4$ coatings which can replace conventional doping while maintaining CMOS-compatibility for ULSI devices \cite{Koe14}.

\begin{figure}[t!]
\begin{center}
\mbox{\scalebox{0.3002}{\includegraphics{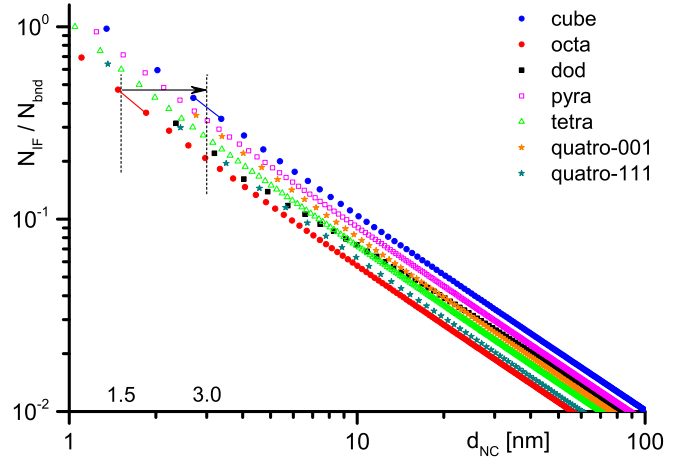}}}
\end{center}
\caption{{\bf Ratio of interface bonds per NC to bonds between atoms of NC $\mathbf{N\mathrm{\mathbf{_{IF}}}/N\mathrm{\mathbf{_{bnd}}}}$.} Size values refer to smallest observed Si NCs and biggest observed size for octahedral Si NCs, see section \ref{Applicat} for further explanation. A linear interpolation between values of the number series  for \{111\}-octahedral and \{001\}-cubic zb-NCs was used to estimate the size range discussed in the text. For legend description see figure \ref{fig10}.}
\label{fig12}
\end{figure}

The quantity $N\mathrm{_{IF}}(d_{\mathrm{NC}}[i])/N_{\mathrm{bnd}}(d_{\mathrm{NC}}[i])$ provides the number of interface bonds per bond between NC atoms, see figure \ref{fig12}. With increasing $d_{\mathrm{NC}}[i]$, the behaviour of $N\mathrm{_{IF}}/N\mathrm{_{bnd}}$ converges against $\frac{1}{2}\,N\mathrm{_{IF}}/N_{\mathrm{NC}}$ due to $\lim_{i\rightarrow\infty}(N\mathrm{_{bnd}}[i]/N_{\mathrm{NC}}[i])=2$, \emph{cf.} figure \ref{fig10}. 

\section{\label{Applicat}Applications}
The ratio $N\mathrm{_{bnd}}/N\mathrm{_{NC}}$ yields information about the NC response to external stress like NC lattice deformation. As example, $N\mathrm{_{bnd}}/N\mathrm{_{NC}}$ is a gauge for the NC stress response to attempts of dopant formation on Si NC lattice sites, triggering self-purification \cite{Steg09,Chan08}. Numerical values of $d_{\mathrm{NC}}$ and $N\mathrm{_{bnd}}/N\mathrm{_{NC}}$ in figure \ref{fig10} refer to figure 7 in \cite{Steg09} where donor formation was shown to decrease from $6\times 10^{-3}$ for 20 nm Si NCs via $1.3\times 10^{-4}$ for 10 nm Si NCs to $3\times 10^{-5}$ for 4.5 nm Si NCs. We can thus derive that self-purification for free-standing Si NCs prevents dopant formation in the NC lattice for $N\mathrm{_{bnd}}/N\mathrm{_{NC}} \leq 1.92\pm 0.02$. We can narrow down this range further if the preferential shape of the NCs was known. Figure 7 in \cite{Steg09} also shows  increased standard deviations of doping efficiencies with shrinking NC size which may originate from an increasing relative size deviation with shrinking NC size. However, we note that $N\mathrm{_{bnd}}/N\mathrm{_{NC}}$ values cover a wider range with shrinking $d_{\mathrm{NC}}$ which may add to the uncertainty in EPR measurements and NC counter-stress related phenomena in general if the NC shape is not known or NC ensembles have no strong preference in shape and faceting. In a similar fashion, $N\mathrm{_{bnd}}/N\mathrm{_{NC}}$ can be useful for investigating dopant clustering in ULSI devices \cite{Kamb13,Koel13} as function of Si nanovolume shape, size and interface orientation.

The ratio $N\mathrm{_{IF}}(d_{\mathrm{NC}})/N_{\mathrm{NC}}(d_{\mathrm{NC}})$ presents a gauge for the impact of interface charge transfer \cite{Koe14,Koe08a} and interface dipoles \cite{Hei65,Ter84,Nis11} onto NCs. Both phenomena have a major influence on NC electronic and optical properties. Interface charge transfer dominates electronic structures of Si NCs for $d_{\mathrm{NC}}\geq 70$ {\AA} \cite{Koe14,Ehrh13}, corresponding values of $N\mathrm{_{IF}}(d_{\mathrm{NC}})/N_{\mathrm{NC}}(d_{\mathrm{NC}})$ are shown in figure \ref{fig11} for \{001\}-faceted cubic and \{111\}-faceted octahedral Si NCs, presenting maxium and minimum values, respectively. Figure \ref{fig11} shows that cubic high symmetry zb-NCs are dominated by the dielectric for $d_{\mathrm{NC}}\leq 12$ nm (dashed blue line). For lower symmetry NCs like Si cubicles of fin-FETs, the size limit of $d_{\mathrm{NC}}$ where the impact of the embedding dielectric dominates the electronic structure of the NC increases further due to an increased ratio $N\mathrm{_{IF}}/N\mathrm{_{NC}}$ \cite{Koe14}.

\begin{figure}[t!]
\begin{center}
\mbox{\scalebox{0.3002}{\includegraphics{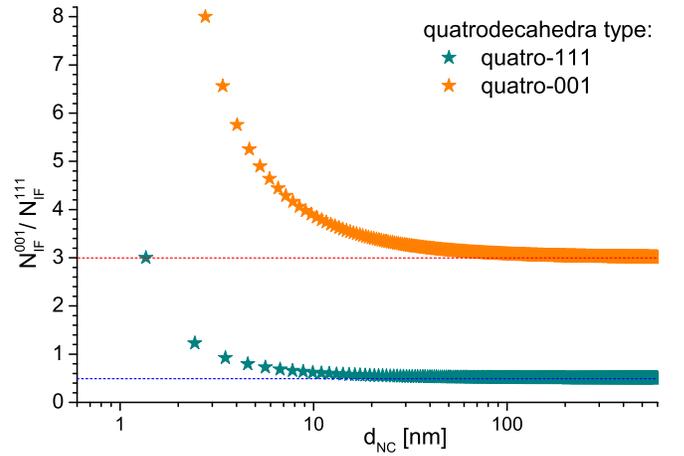}}}
\end{center}
\caption{{\bf Ratio of \{001\} to \{111\} interface bonds for \{111\}- and \{001\}-dominant quadrodecahedral zb-NCs.} Lines show asymptotic values for $d_{\mathrm{NC}}[i]\rightarrow\infty$ of \{111\}-dominated (blue) and \{001\}-dominated quatrodecahedra (red). No overlap exists between both ratios for finite NC size. For legend description see figure \ref{fig10}.}
\label{fig13}
\end{figure}

Partitions of $N\mathrm{_{IF}}(d_{\mathrm{NC}}[i])$ as function of surface orientation can be a cornerstone to quantify EPR-active Si dangling bonds (DBs) at Si NC/SiO$_2$ interfaces \cite{Stes08}. Specific interface defect densities of Si DBs, $P_{b(0)}$ and $P_{b1}$, were assigned to \{001\} and \{111\} planes \cite{Helm94,Keun11}. Si NC shapes were estimated by $P_{b(0)}/P_{b1}$ EPR signal ratios, leading to the postulation of Si NCs \cite{Jiva08} alike to \{001\}-dominated octahedra (section \ref{quatro001}). Partitions of \{001\} and \{111\} planes $N\mathrm{_{IF,\langle 001\rangle}}(d_{\mathrm{NC}}[i])$ and $N\mathrm{_{IF,\langle 111\rangle}}(d_{\mathrm{NC}}[i])$ for \{001\}- and \{111\}-dominated quatrodecahedral Si NCs (equations \ref{eqn-38}, \ref{eqn-39} and \ref{eqn-33}, \ref{eqn-34}, respectively) can be used to interprete EPR data to obtain the dominant NC shape or -- if the dominant NC shape is known -- an estimate for the average NC size. The terms $N\mathrm{_{IF,\langle 001\rangle}^{q.001}}(d_{\mathrm{NC}}[i])/N\mathrm{_{IF,\langle 111\rangle}^{q.001}}(d_{\mathrm{NC}}[i])$ and $N\mathrm{_{IF,\langle 001\rangle}^{q.111}}(d_{\mathrm{NC}}[i])/N\mathrm{_{IF,\langle 111\rangle}^{q.111}}(d_{\mathrm{NC}}[i])$ yield ratios of \{001\}- to \{111\}-oriented interface bonds for which the density of $P_{b(0)}$ and $P_{b1}$ defects are well known from \{001\}- and \{111\}-oriented Si/thermal SiO$_2$ interfaces \cite{Helm94,Keun11}. Figure \ref{fig13} shows these terms as function of $d_{\mathrm{NC}}[i]$. Asymptotic values for infinite NC size are 
$\lim\limits{_{i\rightarrow\infty}}\Big[N\mathrm{_{IF,\langle 001\rangle}^{q.111}}(d_{\mathrm{NC}}[i])/N\mathrm{_{IF,\langle 111\rangle}^{q.111}}(d_{\mathrm{NC}}[i])\Big]=1/2$ for \{111\}-dominated quatrodecahedra and $\lim\limits{_{i\rightarrow\infty}}\Big[N\mathrm{_{IF,\langle 001\rangle}^{q.001}}(d_{\mathrm{NC}}[i])/N\mathrm{_{IF,\langle 111\rangle}^{q.001}}(d_{\mathrm{NC}}[i])\Big]=3$ for \{001\}-dominated quatrodecahedra. No overlap exists between both ratios for finite NC size which allows for a clear assignment of either dominant shape or size of NCs, depending on what parameter is already known.
\\

The quantity $N\mathrm{_{IF}}/N\mathrm{_{bnd}}$ serves as gauge for the stress balance between NCs and an embedding matrix interacting via interface bonds such as Si NCs in SiO$_2$ or strain-balanced growth of InAs QDs in GaNAs \cite{Osh08,Pope08,Bail09}. 
High resolution transmission Electron Microscopy (HR-TEM) was used to determine the size of Si NCs grown by segregation from Si-rich dielectrics. Smallest Si NC extensions were found to be 15 {\AA}, and \{111\}-faceted octahedra dominated the shape of Si NCs up to ca. 30 {\AA} \cite{Gode08}. Disintegration of porous Si by etching and subsequent self-limiting oxidation of Si NCs obtained this way resulted in minimum NC sizes of 12 to 17 {\AA} \cite{Schu94}, the smallest range reported to date for Si NCs obtained by solid-state synthesis. Minimum values of $d_{\mathrm{NC}}=15$ {\AA} were obtained from gas phase synthesis \cite{Steg09}. The minimum diameter of embedded Si NCs should depend on the external stress exerted onto the Si atomic cluster which appears to yield an amorphous structure below $d_{\mathrm{NC}}=15\pm 2$ {\AA}. The predominant existence of embedded Si NCs as \{111\}-faceted octahedra can be explained by their minimum ratio of $N\mathrm{_{IF}}/N\mathrm{_{bnd}}$ which reduces stress exertion. Moreover, Si NC shapes get increasingly polymorphous for sizes beyond 30 {\AA} where $N\mathrm{_{IF}}/N\mathrm{_{bnd}}$ of all other Si NC shapes drops below $N\mathrm{_{IF}^{octa}}/N\mathrm{_{bnd}^{octa}}$ at $d\mathrm{_{NC}^{octa}}=15$ {\AA}, see figure \ref{fig12}. We can therefore derive an upper limit of the bond ratio $N\mathrm{_{IF}}/N\mathrm{_{bnd}}= \frac{1}{476}(197\pm 27)\,\approx  0.414\pm0.057$ below which Si NCs form out of  Si-rich SiO$_2$ or amorphous clusters. For the latter, surface tension appears to be the limiting factor of minimum NC size \cite{Koe06}. Such empirical values of $N\mathrm{_{IF}}/N\mathrm{_{bnd}}$ are a function of the zb solid and its embedding environment and can be derived for any zb NCs with appropriate experimental input. Using material data such as Young's modulus and diffusion coefficients during the NC formation process, we can extrapolate the findings for Si to arrive at estimates for other group IV NCs such as Ge, SiGe and SiC. As an example, the limit of $N\mathrm{_{IF}}/N\mathrm{_{bnd}}$ for Si NCs formed by segregation from Si-rich Si$_3$N$_4$ should be considerably smaller due to the higher Young's modulus and higher packing fraction (hampering diffusion) of Si$_3$N$_4$ as compared to SiO$_2$. As a result, the minimum size of Si NCs formed in Si-rich Si$_3$N$_4$ should be notably bigger as compared to Si NCs formed in Si-rich SiO$_2$.

Stress can be further deconvoluted into NC counter-stress and stress originating from the embedding matrix, with $N\mathrm{_{bnd}}/N\mathrm{_{NC}}$ allowing to quantitatively interpret and deconvolute Raman- and Infrared (IR) spectra. A deconvolution of phonon modes is useful to distinguish between internal (NC) and exteral (embedding matrix) stress as function of NC size, shape and interface orientation \cite{Okad84}. In analogy to Raman- and IR spectroscopy, stress-dependent  photoluminescence (PL) spectra of Si NCs \cite{Kuso12} can be interpreted and deconvoluted. Here, PL or electroluminescence (EL) on single Si NCs are particular useful \cite{Vale02,Vale04} as individual NCs minimize statistical uncertainties due to shape, size and interface termination. Using  $N\mathrm{_{bnd}}/N\mathrm{_{NC}}$ and $N\mathrm{_{IF}}/N\mathrm{_{bnd}}$ for interpreting NC PL spectra as function of applied stress \cite{Iban15} is another method to gain insight into NC properties. 

Ensembles of zb-NCs  will not necessarily have exactly one of the high-symmetry shapes treated in this work, but a rather spherical shape with a mix of different low-index lattice facets defining the NC surface. In particular for NCs grown by segregation processes \cite{Heit05} it is not possible to describe the exact NC shape on a per-NC base due to significant statistical deviations in size and shape \cite{Lau14}. However, with the symmetry arguments derived above, it becomes clear that $N_{\mathrm{NC}}(d_{\mathrm{NC}})$, $N\mathrm{_{bnd}}(d_{\mathrm{NC}})$ and $N\mathrm{_{IF}}(d_{\mathrm{NC}})$ of such NCs are located between values of cubic and octahedral NCs.\\

\section{\label{conclusion} Conclusion}
I deduced analytical number series for zb-NCs as a function of size, shape and surface faceting: the number of NC atoms $N_{\mathrm{NC}}[i]$, the number of bonds existing between such atoms $N\mathrm{_{bnd}}[i]$ and the number of NC interface bonds $N\mathrm{_{IF}}[i]$ for. All expressions are linked to NC sizes $d_{\mathrm{NC}}[i]$ via their run index $i$. As regular shapes with distinct faceting I investigated \{001\} cubes, \{111\} octahedra, \{110\} dodecahedra,
\{001\} (base)/\{111\} (sides) pyramids, \{111\} tetrahedra, \{111\} (dominant)/\{001\} quatrodecahedra, and \{001\} (dominant)/\{111\} quatrodecahedra, see figure \ref{fig01}. 

The ratio $N\mathrm{_{bnd}}/N\mathrm{_{NC}}$ is useful to gauge internal stress of zb-NCs which is key to evaluate NC self-purification and dopant segregation as encountered in impurity doping, and the general stress response of NCs to an external force. Both, $N\mathrm{_{IF}}/N\mathrm{_{bnd}}$ and $N\mathrm{_{bnd}}/N\mathrm{_{NC}}$, can be applied to optical spectroscopy methods such as FT-IR, Raman, PL or EL to interprete and deconvolute spectra into NC-immanent (internal) and matrix (external) components. The ratio $N\mathrm{_{IF}}(d_{\mathrm{NC}})/N_{\mathrm{NC}}(d_{\mathrm{NC}})$ describes the electronic interaction of NCs with the embedding matrix or ligands to gauge the impact of interface dipoles or interface charge transfer onto the NC electronic structure.
Partitions of $N\mathrm{_{IF}}(d_{\mathrm{NC}}[i])$ as function of surface orientation allow to quantify facet-specific interface ratios of EPR-active Si DBs to estimate NC shapes, whereby the postulated existence of \{001\}-dominated quatrodecahedral Si NCs in SiO$_2$ can be verified. Decomposing $N\mathrm{_{IF}^{q.001}}$ into $N\mathrm{_{IF,\langle 001\rangle}^{q.001}}$ and $N\mathrm{_{IF,\langle 111\rangle}^{q.001}}$, specific \{001\} and \{111\} EPR signal ratios of Si DBs are obtained to arrive at much improved assessments of Si NC size and shape. 

The insights into zb-NC structures revealed here allow for major advancements in experimental data interpretation and understanding of III-V, II-VI and diamond-lattice based NCs. For the first time, a general analytical tool exists to quickly assess stress- and interface-related effects in zb-NCs which allows for a deconvolution of experimental data into environment-exerted, interface-related and NC-internal phenomena.

\begin{acknowledgments}
D.K. acknowledges DAAD-Go8 joint research cooperation schemes in 2012, 2014 and 2016 with IMTEK, Freiburg University, Germany. This work was carried out with financial aid from the UNSW Blue Sky research grant 2015.\\
\end{acknowledgments}

\bibliography{GaugingStress-zbNCs_v00_arxiv}

\begin{thebibliography}{49}%
\makeatletter
\providecommand \@ifxundefined [1]{%
 \@ifx{#1\undefined}
}%
\providecommand \@ifnum [1]{%
 \ifnum #1\expandafter \@firstoftwo
 \else \expandafter \@secondoftwo
 \fi
}%
\providecommand \@ifx [1]{%
 \ifx #1\expandafter \@firstoftwo
 \else \expandafter \@secondoftwo
 \fi
}%
\providecommand \natexlab [1]{#1}%
\providecommand \enquote  [1]{``#1''}%
\providecommand \bibnamefont  [1]{#1}%
\providecommand \bibfnamefont [1]{#1}%
\providecommand \citenamefont [1]{#1}%
\providecommand \href@noop [0]{\@secondoftwo}%
\providecommand \href [0]{\begingroup \@sanitize@url \@href}%
\providecommand \@href[1]{\@@startlink{#1}\@@href}%
\providecommand \@@href[1]{\endgroup#1\@@endlink}%
\providecommand \@sanitize@url [0]{\catcode `\\12\catcode `\$12\catcode
  `\&12\catcode `\#12\catcode `\^12\catcode `\_12\catcode `\%12\relax}%
\providecommand \@@startlink[1]{}%
\providecommand \@@endlink[0]{}%
\providecommand \url  [0]{\begingroup\@sanitize@url \@url }%
\providecommand \@url [1]{\endgroup\@href {#1}{\urlprefix }}%
\providecommand \urlprefix  [0]{URL }%
\providecommand \Eprint [0]{\href }%
\providecommand \doibase [0]{http://dx.doi.org/}%
\providecommand \selectlanguage [0]{\@gobble}%
\providecommand \bibinfo  [0]{\@secondoftwo}%
\providecommand \bibfield  [0]{\@secondoftwo}%
\providecommand \translation [1]{[#1]}%
\providecommand \BibitemOpen [0]{}%
\providecommand \bibitemStop [0]{}%
\providecommand \bibitemNoStop [0]{.\EOS\space}%
\providecommand \EOS [0]{\spacefactor3000\relax}%
\providecommand \BibitemShut  [1]{\csname bibitem#1\endcsname}%
\let\auto@bib@innerbib\@empty
\bibitem [{\citenamefont {Jaccodine}\ and\ \citenamefont
  {Schlegel}(1966)}]{Jac66}%
  \BibitemOpen
  \bibfield  {author} {\bibinfo {author} {\bibfnamefont {R.~J.}\ \bibnamefont
  {Jaccodine}}\ and\ \bibinfo {author} {\bibfnamefont {W.~A.}\ \bibnamefont
  {Schlegel}},\ }\href@noop {} {\bibfield  {journal} {\bibinfo  {journal} {J.
  Appl. Phys.}\ }\textbf {\bibinfo {volume} {37}},\ \bibinfo {pages} {2429}
  (\bibinfo {year} {1966})}\BibitemShut {NoStop}%
\bibitem [{\citenamefont {Anastassakis}\ \emph {et~al.}(1970)\citenamefont
  {Anastassakis}, \citenamefont {Pinczuk}, \citenamefont {Burstein},
  \citenamefont {Pollack},\ and\ \citenamefont {Cardona}}]{Ana70}%
  \BibitemOpen
  \bibfield  {author} {\bibinfo {author} {\bibfnamefont {E.}~\bibnamefont
  {Anastassakis}}, \bibinfo {author} {\bibfnamefont {A.}~\bibnamefont
  {Pinczuk}}, \bibinfo {author} {\bibfnamefont {E.}~\bibnamefont {Burstein}},
  \bibinfo {author} {\bibfnamefont {F.~H.}\ \bibnamefont {Pollack}}, \ and\
  \bibinfo {author} {\bibfnamefont {M.}~\bibnamefont {Cardona}},\ }\href@noop
  {} {\bibfield  {journal} {\bibinfo  {journal} {Sol. Stat. Comm.}\ }\textbf
  {\bibinfo {volume} {8}},\ \bibinfo {pages} {133} (\bibinfo {year}
  {1970})}\BibitemShut {NoStop}%
\bibitem [{\citenamefont {Nakashima}\ \emph {et~al.}(1981)\citenamefont
  {Nakashima}, \citenamefont {Oima}, \citenamefont {Mitsuishi}, \citenamefont
  {Nishimura}, \citenamefont {Fukumoto},\ and\ \citenamefont
  {Akasaka}}]{Nak81}%
  \BibitemOpen
  \bibfield  {author} {\bibinfo {author} {\bibfnamefont {S.}~\bibnamefont
  {Nakashima}}, \bibinfo {author} {\bibfnamefont {S.}~\bibnamefont {Oima}},
  \bibinfo {author} {\bibfnamefont {A.}~\bibnamefont {Mitsuishi}}, \bibinfo
  {author} {\bibfnamefont {T.}~\bibnamefont {Nishimura}}, \bibinfo {author}
  {\bibfnamefont {T.}~\bibnamefont {Fukumoto}}, \ and\ \bibinfo {author}
  {\bibfnamefont {Y.}~\bibnamefont {Akasaka}},\ }\href@noop {} {\bibfield
  {journal} {\bibinfo  {journal} {Sol. Stat. Comm.}\ }\textbf {\bibinfo
  {volume} {40}},\ \bibinfo {pages} {765} (\bibinfo {year} {1981})}\BibitemShut
  {NoStop}%
\bibitem [{\citenamefont {Boyd}\ and\ \citenamefont {Wilson}(1982)}]{Boy82}%
  \BibitemOpen
  \bibfield  {author} {\bibinfo {author} {\bibfnamefont {I.~W.}\ \bibnamefont
  {Boyd}}\ and\ \bibinfo {author} {\bibfnamefont {J.~I.~B.}\ \bibnamefont
  {Wilson}},\ }\href@noop {} {\bibfield  {journal} {\bibinfo  {journal} {J.
  Appl. Phys.}\ }\textbf {\bibinfo {volume} {53}},\ \bibinfo {pages} {4166}
  (\bibinfo {year} {1982})}\BibitemShut {NoStop}%
\bibitem [{\citenamefont {Boyd}\ and\ \citenamefont {Wilson}(1987)}]{Boy87}%
  \BibitemOpen
  \bibfield  {author} {\bibinfo {author} {\bibfnamefont {I.~W.}\ \bibnamefont
  {Boyd}}\ and\ \bibinfo {author} {\bibfnamefont {J.~I.~B.}\ \bibnamefont
  {Wilson}},\ }\href@noop {} {\bibfield  {journal} {\bibinfo  {journal} {J.
  Appl. Phys.}\ }\textbf {\bibinfo {volume} {62}},\ \bibinfo {pages} {3195}
  (\bibinfo {year} {1987})}\BibitemShut {NoStop}%
\bibitem [{\citenamefont {Elliot}(1998)}]{Ell98}%
  \BibitemOpen
  \bibfield  {author} {\bibinfo {author} {\bibfnamefont {S.~R.}\ \bibnamefont
  {Elliot}},\ }\href@noop {} {\emph {\bibinfo {title} {The Physics and
  Chemistry of Solids}}}\ (\bibinfo  {publisher} {Wiley},\ \bibinfo {year}
  {1998})\BibitemShut {NoStop}%
\bibitem [{\citenamefont {Smith}\ \emph {et~al.}(2009)\citenamefont {Smith},
  \citenamefont {Mohs},\ and\ \citenamefont {Nie}}]{Smi09}%
  \BibitemOpen
  \bibfield  {author} {\bibinfo {author} {\bibfnamefont {A.~M.}\ \bibnamefont
  {Smith}}, \bibinfo {author} {\bibfnamefont {A.~M.}\ \bibnamefont {Mohs}}, \
  and\ \bibinfo {author} {\bibfnamefont {S.}~\bibnamefont {Nie}},\ }\href@noop
  {} {\bibfield  {journal} {\bibinfo  {journal} {Nature Nanotech.}\ }\textbf
  {\bibinfo {volume} {4}},\ \bibinfo {pages} {56} (\bibinfo {year}
  {2009})}\BibitemShut {NoStop}%
\bibitem [{\citenamefont {Klime\v{s}ov\'{a}}\ \emph {et~al.}(2012)\citenamefont
  {Klime\v{s}ov\'{a}}, \citenamefont {K\r{u}sov\'{a}}, \citenamefont
  {Vac\'{i}k}, \citenamefont {Hol\'{y}},\ and\ \citenamefont {Pelant}}]{Cli12}%
  \BibitemOpen
  \bibfield  {author} {\bibinfo {author} {\bibfnamefont {E.}~\bibnamefont
  {Klime\v{s}ov\'{a}}}, \bibinfo {author} {\bibfnamefont {K.}~\bibnamefont
  {K\r{u}sov\'{a}}}, \bibinfo {author} {\bibfnamefont {J.}~\bibnamefont
  {Vac\'{i}k}}, \bibinfo {author} {\bibfnamefont {V.}~\bibnamefont {Hol\'{y}}},
  \ and\ \bibinfo {author} {\bibfnamefont {I.}~\bibnamefont {Pelant}},\
  }\href@noop {} {\bibfield  {journal} {\bibinfo  {journal} {J. Appl. Phys.}\
  }\textbf {\bibinfo {volume} {112}},\ \bibinfo {pages} {064322} (\bibinfo
  {year} {2012})}\BibitemShut {NoStop}%
\bibitem [{\citenamefont {Laube}\ \emph {et~al.}(2014)\citenamefont {Laube},
  \citenamefont {Gutsch}, \citenamefont {Hiller}, \citenamefont {Bruns},
  \citenamefont {K\"{u}bel}, \citenamefont {Weiss},\ and\ \citenamefont
  {Zacharias}}]{Lau14}%
  \BibitemOpen
  \bibfield  {author} {\bibinfo {author} {\bibfnamefont {J.}~\bibnamefont
  {Laube}}, \bibinfo {author} {\bibfnamefont {S.}~\bibnamefont {Gutsch}},
  \bibinfo {author} {\bibfnamefont {D.}~\bibnamefont {Hiller}}, \bibinfo
  {author} {\bibfnamefont {M.}~\bibnamefont {Bruns}}, \bibinfo {author}
  {\bibfnamefont {C.}~\bibnamefont {K\"{u}bel}}, \bibinfo {author}
  {\bibfnamefont {C.}~\bibnamefont {Weiss}}, \ and\ \bibinfo {author}
  {\bibfnamefont {M.}~\bibnamefont {Zacharias}},\ }\href@noop {} {\bibfield
  {journal} {\bibinfo  {journal} {J. Appl. Phys.}\ }\textbf {\bibinfo {volume}
  {116}},\ \bibinfo {pages} {223501} (\bibinfo {year} {2014})}\BibitemShut
  {NoStop}%
\bibitem [{\citenamefont {Stranski}\ and\ \citenamefont
  {Krastanow}(1938)}]{StraKras38}%
  \BibitemOpen
  \bibfield  {author} {\bibinfo {author} {\bibfnamefont {I.~N.}\ \bibnamefont
  {Stranski}}\ and\ \bibinfo {author} {\bibfnamefont {L.}~\bibnamefont
  {Krastanow}},\ }\href@noop {} {\bibfield  {journal} {\bibinfo  {journal}
  {{Abhandlungen der Mathematisch-Naturwissenschaftlichen Klasse IIb. Akademie
  der Wissenschaften Wien (in German)}}\ }\textbf {\bibinfo {volume} {146}},\
  \bibinfo {pages} {797} (\bibinfo {year} {1938})}\BibitemShut {NoStop}%
\bibitem [{\citenamefont {Bauer}(1958)}]{Bau58}%
  \BibitemOpen
  \bibfield  {author} {\bibinfo {author} {\bibfnamefont {E.}~\bibnamefont
  {Bauer}},\ }\href@noop {} {\bibfield  {journal} {\bibinfo  {journal}
  {Zeitschrift f\"{u}r Kristallographie (in German)}\ }\textbf {\bibinfo
  {volume} {110}},\ \bibinfo {pages} {372} (\bibinfo {year}
  {1958})}\BibitemShut {NoStop}%
\bibitem [{\citenamefont {Oshima}\ \emph {et~al.}(2006)\citenamefont {Oshima},
  \citenamefont {Hashimoto}, \citenamefont {Shigekawa},\ and\ \citenamefont
  {Okada}}]{Osh06}%
  \BibitemOpen
  \bibfield  {author} {\bibinfo {author} {\bibfnamefont {R.}~\bibnamefont
  {Oshima}}, \bibinfo {author} {\bibfnamefont {T.}~\bibnamefont {Hashimoto}},
  \bibinfo {author} {\bibfnamefont {H.}~\bibnamefont {Shigekawa}}, \ and\
  \bibinfo {author} {\bibfnamefont {Y.}~\bibnamefont {Okada}},\ }\href@noop {}
  {\bibfield  {journal} {\bibinfo  {journal} {J. Appl. Phys.}\ }\textbf
  {\bibinfo {volume} {100}},\ \bibinfo {pages} {083110} (\bibinfo {year}
  {2006})}\BibitemShut {NoStop}%
\bibitem [{\citenamefont {Oshima}\ \emph {et~al.}(2008)\citenamefont {Oshima},
  \citenamefont {Takata},\ and\ \citenamefont {Okada}}]{Osh08}%
  \BibitemOpen
  \bibfield  {author} {\bibinfo {author} {\bibfnamefont {R.}~\bibnamefont
  {Oshima}}, \bibinfo {author} {\bibfnamefont {A.}~\bibnamefont {Takata}}, \
  and\ \bibinfo {author} {\bibfnamefont {Y.}~\bibnamefont {Okada}},\
  }\href@noop {} {\bibfield  {journal} {\bibinfo  {journal} {Appl. Phys.
  Lett.}\ }\textbf {\bibinfo {volume} {93}},\ \bibinfo {pages} {083111}
  (\bibinfo {year} {2008})}\BibitemShut {NoStop}%
\bibitem [{\citenamefont {Popescu}\ \emph {et~al.}(2008)\citenamefont
  {Popescu}, \citenamefont {Bester}, \citenamefont {Hanna}, \citenamefont
  {Norman},\ and\ \citenamefont {Zunger}}]{Pope08}%
  \BibitemOpen
  \bibfield  {author} {\bibinfo {author} {\bibfnamefont {V.}~\bibnamefont
  {Popescu}}, \bibinfo {author} {\bibfnamefont {G.}~\bibnamefont {Bester}},
  \bibinfo {author} {\bibfnamefont {M.~C.}\ \bibnamefont {Hanna}}, \bibinfo
  {author} {\bibfnamefont {A.~G.}\ \bibnamefont {Norman}}, \ and\ \bibinfo
  {author} {\bibfnamefont {A.}~\bibnamefont {Zunger}},\ }\href@noop {}
  {\bibfield  {journal} {\bibinfo  {journal} {Phys. Rev. B}\ }\textbf {\bibinfo
  {volume} {78}},\ \bibinfo {pages} {205321} (\bibinfo {year}
  {2008})}\BibitemShut {NoStop}%
\bibitem [{\citenamefont {Bailey}\ \emph {et~al.}(2009)\citenamefont {Bailey},
  \citenamefont {Hubbard}, \citenamefont {Forbes},\ and\ \citenamefont
  {Rafaelle}}]{Bail09}%
  \BibitemOpen
  \bibfield  {author} {\bibinfo {author} {\bibfnamefont {C.~G.}\ \bibnamefont
  {Bailey}}, \bibinfo {author} {\bibfnamefont {S.~M.}\ \bibnamefont {Hubbard}},
  \bibinfo {author} {\bibfnamefont {D.~V.}\ \bibnamefont {Forbes}}, \ and\
  \bibinfo {author} {\bibfnamefont {R.~P.}\ \bibnamefont {Rafaelle}},\
  }\href@noop {} {\bibfield  {journal} {\bibinfo  {journal} {Appl. Phys.
  Lett.}\ }\textbf {\bibinfo {volume} {95}},\ \bibinfo {pages} {203110}
  (\bibinfo {year} {2009})}\BibitemShut {NoStop}%
\bibitem [{\citenamefont {Stegner}\ \emph {et~al.}(2008)\citenamefont
  {Stegner}, \citenamefont {Pereira}, \citenamefont {Klein}, \citenamefont
  {Lechner}, \citenamefont {Dietmueller}, \citenamefont {Brandt}, \citenamefont
  {Stutzmann},\ and\ \citenamefont {Wiggers}}]{Steg08a}%
  \BibitemOpen
  \bibfield  {author} {\bibinfo {author} {\bibfnamefont {A.~R.}\ \bibnamefont
  {Stegner}}, \bibinfo {author} {\bibfnamefont {R.~N.}\ \bibnamefont
  {Pereira}}, \bibinfo {author} {\bibfnamefont {K.}~\bibnamefont {Klein}},
  \bibinfo {author} {\bibfnamefont {R.}~\bibnamefont {Lechner}}, \bibinfo
  {author} {\bibfnamefont {R.}~\bibnamefont {Dietmueller}}, \bibinfo {author}
  {\bibfnamefont {M.~S.}\ \bibnamefont {Brandt}}, \bibinfo {author}
  {\bibfnamefont {M.}~\bibnamefont {Stutzmann}}, \ and\ \bibinfo {author}
  {\bibfnamefont {H.}~\bibnamefont {Wiggers}},\ }\href@noop {} {\bibfield
  {journal} {\bibinfo  {journal} {Phys. Rev. Lett.}\ }\textbf {\bibinfo
  {volume} {100}},\ \bibinfo {pages} {026803} (\bibinfo {year}
  {2008})}\BibitemShut {NoStop}%
\bibitem [{\citenamefont {Stegner}\ \emph {et~al.}(2009)\citenamefont
  {Stegner}, \citenamefont {Pereira}, \citenamefont {Lechner}, \citenamefont
  {Klein}, \citenamefont {Wiggers}, \citenamefont {Stutzmann},\ and\
  \citenamefont {Brandt}}]{Steg09}%
  \BibitemOpen
  \bibfield  {author} {\bibinfo {author} {\bibfnamefont {A.~R.}\ \bibnamefont
  {Stegner}}, \bibinfo {author} {\bibfnamefont {R.~N.}\ \bibnamefont
  {Pereira}}, \bibinfo {author} {\bibfnamefont {R.}~\bibnamefont {Lechner}},
  \bibinfo {author} {\bibfnamefont {K.}~\bibnamefont {Klein}}, \bibinfo
  {author} {\bibfnamefont {H.}~\bibnamefont {Wiggers}}, \bibinfo {author}
  {\bibfnamefont {M.}~\bibnamefont {Stutzmann}}, \ and\ \bibinfo {author}
  {\bibfnamefont {M.~S.}\ \bibnamefont {Brandt}},\ }\href@noop {} {\bibfield
  {journal} {\bibinfo  {journal} {Phys. Rev. B}\ }\textbf {\bibinfo {volume}
  {80}},\ \bibinfo {pages} {165326} (\bibinfo {year} {2009})}\BibitemShut
  {NoStop}%
\bibitem [{\citenamefont {K\"onig}\ \emph {et~al.}(2015)\citenamefont
  {K\"onig}, \citenamefont {Gutsch}, \citenamefont {Gnaser}, \citenamefont
  {Kopnarski}, \citenamefont {G{\"o}ttlicher}, \citenamefont {Steininger},
  \citenamefont {Zacharias},\ and\ \citenamefont {Hiller}}]{Koe15a}%
  \BibitemOpen
  \bibfield  {author} {\bibinfo {author} {\bibfnamefont {D.}~\bibnamefont
  {K\"onig}}, \bibinfo {author} {\bibfnamefont {S.}~\bibnamefont {Gutsch}},
  \bibinfo {author} {\bibfnamefont {H.}~\bibnamefont {Gnaser}}, \bibinfo
  {author} {\bibfnamefont {M.}~\bibnamefont {Kopnarski}}, \bibinfo {author}
  {\bibfnamefont {J.}~\bibnamefont {G{\"o}ttlicher}}, \bibinfo {author}
  {\bibfnamefont {R.}~\bibnamefont {Steininger}}, \bibinfo {author}
  {\bibfnamefont {M.}~\bibnamefont {Zacharias}}, \ and\ \bibinfo {author}
  {\bibfnamefont {D.}~\bibnamefont {Hiller}},\ }\href@noop {} {\bibfield
  {journal} {\bibinfo  {journal} {Sci. Rep.}\ }\textbf {\bibinfo {volume}
  {5}},\ \bibinfo {pages} {9702} (\bibinfo {year} {2015})}\BibitemShut
  {NoStop}%
\bibitem [{\citenamefont {Gnaser}\ \emph {et~al.}(2014)\citenamefont {Gnaser},
  \citenamefont {Gutsch}, \citenamefont {Wahl}, \citenamefont {Schiller},
  \citenamefont {Kopnarski}, \citenamefont {Hiller},\ and\ \citenamefont
  {Zacharias}}]{Gnas14}%
  \BibitemOpen
  \bibfield  {author} {\bibinfo {author} {\bibfnamefont {H.}~\bibnamefont
  {Gnaser}}, \bibinfo {author} {\bibfnamefont {S.}~\bibnamefont {Gutsch}},
  \bibinfo {author} {\bibfnamefont {M.}~\bibnamefont {Wahl}}, \bibinfo {author}
  {\bibfnamefont {R.}~\bibnamefont {Schiller}}, \bibinfo {author}
  {\bibfnamefont {M.}~\bibnamefont {Kopnarski}}, \bibinfo {author}
  {\bibfnamefont {D.}~\bibnamefont {Hiller}}, \ and\ \bibinfo {author}
  {\bibfnamefont {M.}~\bibnamefont {Zacharias}},\ }\href@noop {} {\bibfield
  {journal} {\bibinfo  {journal} {J. Appl. Phys.}\ }\textbf {\bibinfo {volume}
  {115}},\ \bibinfo {pages} {034304} (\bibinfo {year} {2014})}\BibitemShut
  {NoStop}%
\bibitem [{\citenamefont {Dalpian}\ and\ \citenamefont
  {Chelikowsky}(2006)}]{Dalp06}%
  \BibitemOpen
  \bibfield  {author} {\bibinfo {author} {\bibfnamefont {G.~M.}\ \bibnamefont
  {Dalpian}}\ and\ \bibinfo {author} {\bibfnamefont {J.~R.}\ \bibnamefont
  {Chelikowsky}},\ }\href@noop {} {\bibfield  {journal} {\bibinfo  {journal}
  {Phys. Rev. Lett.}\ }\textbf {\bibinfo {volume} {96}},\ \bibinfo {pages}
  {226802} (\bibinfo {year} {2006})}\BibitemShut {NoStop}%
\bibitem [{\citenamefont {Dalpian}\ and\ \citenamefont
  {Chelikowsky}(2008)}]{Dalp08}%
  \BibitemOpen
  \bibfield  {author} {\bibinfo {author} {\bibfnamefont {G.~M.}\ \bibnamefont
  {Dalpian}}\ and\ \bibinfo {author} {\bibfnamefont {J.~R.}\ \bibnamefont
  {Chelikowsky}},\ }\href@noop {} {\bibfield  {journal} {\bibinfo  {journal}
  {Phys. Rev. Lett.}\ }\textbf {\bibinfo {volume} {100}},\ \bibinfo {pages}
  {179703} (\bibinfo {year} {2008})}\BibitemShut {NoStop}%
\bibitem [{\citenamefont {Chan}\ \emph {et~al.}(2008)\citenamefont {Chan},
  \citenamefont {Tiago}, \citenamefont {Kaxiras},\ and\ \citenamefont
  {Chelikowsky}}]{Chan08}%
  \BibitemOpen
  \bibfield  {author} {\bibinfo {author} {\bibfnamefont {T.~L.}\ \bibnamefont
  {Chan}}, \bibinfo {author} {\bibfnamefont {M.~L.}\ \bibnamefont {Tiago}},
  \bibinfo {author} {\bibfnamefont {E.}~\bibnamefont {Kaxiras}}, \ and\
  \bibinfo {author} {\bibfnamefont {J.~R.}\ \bibnamefont {Chelikowsky}},\
  }\href@noop {} {\bibfield  {journal} {\bibinfo  {journal} {Nano Lett.}\
  }\textbf {\bibinfo {volume} {8}},\ \bibinfo {pages} {596} (\bibinfo {year}
  {2008})}\BibitemShut {NoStop}%
\bibitem [{\citenamefont {Ossicini}\ \emph {et~al.}(2005)\citenamefont
  {Ossicini}, \citenamefont {Degoli}, \citenamefont {Iori}, \citenamefont
  {Luppi}, \citenamefont {Magri}, \citenamefont {Cantele}, \citenamefont
  {Trani},\ and\ \citenamefont {Ninno}}]{Ossi05}%
  \BibitemOpen
  \bibfield  {author} {\bibinfo {author} {\bibfnamefont {S.}~\bibnamefont
  {Ossicini}}, \bibinfo {author} {\bibfnamefont {E.}~\bibnamefont {Degoli}},
  \bibinfo {author} {\bibfnamefont {F.}~\bibnamefont {Iori}}, \bibinfo {author}
  {\bibfnamefont {E.}~\bibnamefont {Luppi}}, \bibinfo {author} {\bibfnamefont
  {R.}~\bibnamefont {Magri}}, \bibinfo {author} {\bibfnamefont
  {G.}~\bibnamefont {Cantele}}, \bibinfo {author} {\bibfnamefont
  {F.}~\bibnamefont {Trani}}, \ and\ \bibinfo {author} {\bibfnamefont
  {D.}~\bibnamefont {Ninno}},\ }\href@noop {} {\bibfield  {journal} {\bibinfo
  {journal} {Appl. Phys. Lett.}\ }\textbf {\bibinfo {volume} {87}},\ \bibinfo
  {pages} {173120} (\bibinfo {year} {2005})}\BibitemShut {NoStop}%
\bibitem [{\citenamefont {K{\"o}nig}\ \emph {et~al.}(2014)\citenamefont
  {K{\"o}nig}, \citenamefont {Hiller}, \citenamefont {Gutsch},\ and\
  \citenamefont {Zacharias}}]{Koe14}%
  \BibitemOpen
  \bibfield  {author} {\bibinfo {author} {\bibfnamefont {D.}~\bibnamefont
  {K{\"o}nig}}, \bibinfo {author} {\bibfnamefont {D.}~\bibnamefont {Hiller}},
  \bibinfo {author} {\bibfnamefont {S.}~\bibnamefont {Gutsch}}, \ and\ \bibinfo
  {author} {\bibfnamefont {M.}~\bibnamefont {Zacharias}},\ }\href@noop {}
  {\bibfield  {journal} {\bibinfo  {journal} {Appl. Mater. Interfaces}\
  }\textbf {\bibinfo {volume} {1}},\ \bibinfo {pages} {1400359} (\bibinfo
  {year} {2014})}\BibitemShut {NoStop}%
\bibitem [{\citenamefont {Hesketh}\ \emph {et~al.}(1993)\citenamefont
  {Hesketh}, \citenamefont {Ju}, \citenamefont {Gowda}, \citenamefont
  {Zanoria},\ and\ \citenamefont {Danyluk}}]{Hes93}%
  \BibitemOpen
  \bibfield  {author} {\bibinfo {author} {\bibfnamefont {P.~J.}\ \bibnamefont
  {Hesketh}}, \bibinfo {author} {\bibfnamefont {C.}~\bibnamefont {Ju}},
  \bibinfo {author} {\bibfnamefont {S.}~\bibnamefont {Gowda}}, \bibinfo
  {author} {\bibfnamefont {E.}~\bibnamefont {Zanoria}}, \ and\ \bibinfo
  {author} {\bibfnamefont {S.}~\bibnamefont {Danyluk}},\ }\href@noop {}
  {\bibfield  {journal} {\bibinfo  {journal} {J. Electrochem. Soc.}\ }\textbf
  {\bibinfo {volume} {140}},\ \bibinfo {pages} {1080} (\bibinfo {year}
  {1993})}\BibitemShut {NoStop}%
\bibitem [{\citenamefont {Eaglesham}\ \emph {et~al.}(1993)\citenamefont
  {Eaglesham}, \citenamefont {White}, \citenamefont {Feldman}, \citenamefont
  {Moriya},\ and\ \citenamefont {Jacobson}}]{Eagl93}%
  \BibitemOpen
  \bibfield  {author} {\bibinfo {author} {\bibfnamefont {D.~J.}\ \bibnamefont
  {Eaglesham}}, \bibinfo {author} {\bibfnamefont {A.~E.}\ \bibnamefont
  {White}}, \bibinfo {author} {\bibfnamefont {L.~C.}\ \bibnamefont {Feldman}},
  \bibinfo {author} {\bibfnamefont {N.}~\bibnamefont {Moriya}}, \ and\ \bibinfo
  {author} {\bibfnamefont {D.~C.}\ \bibnamefont {Jacobson}},\ }\href@noop {}
  {\bibfield  {journal} {\bibinfo  {journal} {Phys. Rev. Lett.}\ }\textbf
  {\bibinfo {volume} {70}},\ \bibinfo {pages} {1643} (\bibinfo {year}
  {1993})}\BibitemShut {NoStop}%
\bibitem [{\citenamefont {Godefroo}\ \emph {et~al.}(2008)\citenamefont
  {Godefroo}, \citenamefont {Hayne}, \citenamefont {Jivanescu}, \citenamefont
  {Stesmans}, \citenamefont {Zacharias}, \citenamefont {Lebedev}, \citenamefont
  {Tendeloo},\ and\ \citenamefont {Moshchalkov}}]{Gode08}%
  \BibitemOpen
  \bibfield  {author} {\bibinfo {author} {\bibfnamefont {S.}~\bibnamefont
  {Godefroo}}, \bibinfo {author} {\bibfnamefont {M.}~\bibnamefont {Hayne}},
  \bibinfo {author} {\bibfnamefont {M.}~\bibnamefont {Jivanescu}}, \bibinfo
  {author} {\bibfnamefont {A.}~\bibnamefont {Stesmans}}, \bibinfo {author}
  {\bibfnamefont {M.}~\bibnamefont {Zacharias}}, \bibinfo {author}
  {\bibfnamefont {O.~I.}\ \bibnamefont {Lebedev}}, \bibinfo {author}
  {\bibfnamefont {G.~V.}\ \bibnamefont {Tendeloo}}, \ and\ \bibinfo {author}
  {\bibfnamefont {V.~V.}\ \bibnamefont {Moshchalkov}},\ }\href@noop {}
  {\bibfield  {journal} {\bibinfo  {journal} {Nature Nanotech.}\ }\textbf
  {\bibinfo {volume} {3}},\ \bibinfo {pages} {174} (\bibinfo {year}
  {2008})}\BibitemShut {NoStop}%
\bibitem [{\citenamefont {Zeidler}(2004)}]{Zeid04}%
  \BibitemOpen
  \bibinfo {editor} {\bibfnamefont {E.}~\bibnamefont {Zeidler}},\ ed.,\
  \href@noop {} {\emph {\bibinfo {title} {Oxford Users' Guide to Mathematics
  (translated from German by B. Hunt)}}}\ (\bibinfo  {publisher} {Oxford
  University Press},\ \bibinfo {year} {2004})\BibitemShut {NoStop}%
\bibitem [{\citenamefont {Grundmann}\ \emph {et~al.}(1995)\citenamefont
  {Grundmann}, \citenamefont {Stier},\ and\ \citenamefont {Bimberg}}]{Grun95}%
  \BibitemOpen
  \bibfield  {author} {\bibinfo {author} {\bibfnamefont {M.}~\bibnamefont
  {Grundmann}}, \bibinfo {author} {\bibfnamefont {O.}~\bibnamefont {Stier}}, \
  and\ \bibinfo {author} {\bibfnamefont {D.}~\bibnamefont {Bimberg}},\
  }\href@noop {} {\bibfield  {journal} {\bibinfo  {journal} {Phys. Rev. B}\
  }\textbf {\bibinfo {volume} {52}},\ \bibinfo {pages} {11969} (\bibinfo {year}
  {1995})}\BibitemShut {NoStop}%
\bibitem [{\citenamefont {Baier}\ \emph {et~al.}(2004)\citenamefont {Baier},
  \citenamefont {Pelucchi}, \citenamefont {Kapon}, \citenamefont {Varoutsis},
  \citenamefont {Gallart}, \citenamefont {Robert-Philip},\ and\ \citenamefont
  {Abram}}]{Bai04}%
  \BibitemOpen
  \bibfield  {author} {\bibinfo {author} {\bibfnamefont {M.~H.}\ \bibnamefont
  {Baier}}, \bibinfo {author} {\bibfnamefont {E.}~\bibnamefont {Pelucchi}},
  \bibinfo {author} {\bibfnamefont {E.}~\bibnamefont {Kapon}}, \bibinfo
  {author} {\bibfnamefont {S.}~\bibnamefont {Varoutsis}}, \bibinfo {author}
  {\bibfnamefont {M.}~\bibnamefont {Gallart}}, \bibinfo {author} {\bibfnamefont
  {I.}~\bibnamefont {Robert-Philip}}, \ and\ \bibinfo {author} {\bibfnamefont
  {I.}~\bibnamefont {Abram}},\ }\href@noop {} {\bibfield  {journal} {\bibinfo
  {journal} {Appl. Phys. Lett.}\ }\textbf {\bibinfo {volume} {84}},\ \bibinfo
  {pages} {648} (\bibinfo {year} {2004})}\BibitemShut {NoStop}%
\bibitem [{\citenamefont {Stesmans}\ \emph {et~al.}(2008)\citenamefont
  {Stesmans}, \citenamefont {Jivanescu}, \citenamefont {Godefroo},\ and\
  \citenamefont {Zacharias}}]{Stes08}%
  \BibitemOpen
  \bibfield  {author} {\bibinfo {author} {\bibfnamefont {A.}~\bibnamefont
  {Stesmans}}, \bibinfo {author} {\bibfnamefont {M.}~\bibnamefont {Jivanescu}},
  \bibinfo {author} {\bibfnamefont {S.}~\bibnamefont {Godefroo}}, \ and\
  \bibinfo {author} {\bibfnamefont {M.}~\bibnamefont {Zacharias}},\ }\href@noop
  {} {\bibfield  {journal} {\bibinfo  {journal} {Appl. Phys. Lett.}\ }\textbf
  {\bibinfo {volume} {93}},\ \bibinfo {pages} {023123} (\bibinfo {year}
  {2008})}\BibitemShut {NoStop}%
\bibitem [{\citenamefont {Kambham}\ \emph {et~al.}(2013)\citenamefont
  {Kambham}, \citenamefont {Kumar}, \citenamefont {Florakis},\ and\
  \citenamefont {Vandervorst}}]{Kamb13}%
  \BibitemOpen
  \bibfield  {author} {\bibinfo {author} {\bibfnamefont {A.~K.}\ \bibnamefont
  {Kambham}}, \bibinfo {author} {\bibfnamefont {A.}~\bibnamefont {Kumar}},
  \bibinfo {author} {\bibfnamefont {A.}~\bibnamefont {Florakis}}, \ and\
  \bibinfo {author} {\bibfnamefont {W.}~\bibnamefont {Vandervorst}},\
  }\href@noop {} {\bibfield  {journal} {\bibinfo  {journal} {Nanotechnology}\
  }\textbf {\bibinfo {volume} {24}},\ \bibinfo {pages} {275705} (\bibinfo
  {year} {2013})}\BibitemShut {NoStop}%
\bibitem [{\citenamefont {Koelling}\ \emph {et~al.}(2013)\citenamefont
  {Koelling}, \citenamefont {Richard}, \citenamefont {Bender}, \citenamefont
  {Uematsu}, \citenamefont {Schulze}, \citenamefont {Zschaetzsch},
  \citenamefont {Gilbert},\ and\ \citenamefont {Vandervorst}}]{Koel13}%
  \BibitemOpen
  \bibfield  {author} {\bibinfo {author} {\bibfnamefont {S.}~\bibnamefont
  {Koelling}}, \bibinfo {author} {\bibfnamefont {O.}~\bibnamefont {Richard}},
  \bibinfo {author} {\bibfnamefont {H.}~\bibnamefont {Bender}}, \bibinfo
  {author} {\bibfnamefont {M.}~\bibnamefont {Uematsu}}, \bibinfo {author}
  {\bibfnamefont {A.}~\bibnamefont {Schulze}}, \bibinfo {author} {\bibfnamefont
  {G.}~\bibnamefont {Zschaetzsch}}, \bibinfo {author} {\bibfnamefont
  {M.}~\bibnamefont {Gilbert}}, \ and\ \bibinfo {author} {\bibfnamefont
  {W.}~\bibnamefont {Vandervorst}},\ }\href@noop {} {\bibfield  {journal}
  {\bibinfo  {journal} {Nano Lett.}\ }\textbf {\bibinfo {volume} {13}},\
  \bibinfo {pages} {2458} (\bibinfo {year} {2013})}\BibitemShut {NoStop}%
\bibitem [{\citenamefont {K{\"o}nig}\ \emph {et~al.}(2008)\citenamefont
  {K{\"o}nig}, \citenamefont {Rudd}, \citenamefont {Green},\ and\ \citenamefont
  {Conibeer}}]{Koe08a}%
  \BibitemOpen
  \bibfield  {author} {\bibinfo {author} {\bibfnamefont {D.}~\bibnamefont
  {K{\"o}nig}}, \bibinfo {author} {\bibfnamefont {J.}~\bibnamefont {Rudd}},
  \bibinfo {author} {\bibfnamefont {M.~A.}\ \bibnamefont {Green}}, \ and\
  \bibinfo {author} {\bibfnamefont {G.}~\bibnamefont {Conibeer}},\ }\href@noop
  {} {\bibfield  {journal} {\bibinfo  {journal} {Phys. Rev. B}\ }\textbf
  {\bibinfo {volume} {78}},\ \bibinfo {pages} {035339} (\bibinfo {year}
  {2008})}\BibitemShut {NoStop}%
\bibitem [{\citenamefont {Heine}(1965)}]{Hei65}%
  \BibitemOpen
  \bibfield  {author} {\bibinfo {author} {\bibfnamefont {V.}~\bibnamefont
  {Heine}},\ }\href@noop {} {\bibfield  {journal} {\bibinfo  {journal} {Phys.
  Rev.}\ }\textbf {\bibinfo {volume} {138}},\ \bibinfo {pages} {A1689}
  (\bibinfo {year} {1965})}\BibitemShut {NoStop}%
\bibitem [{\citenamefont {Tersoff}(1984)}]{Ter84}%
  \BibitemOpen
  \bibfield  {author} {\bibinfo {author} {\bibfnamefont {J.}~\bibnamefont
  {Tersoff}},\ }\href@noop {} {\bibfield  {journal} {\bibinfo  {journal} {Phys.
  Rev. B}\ }\textbf {\bibinfo {volume} {30}},\ \bibinfo {pages} {4874}
  (\bibinfo {year} {1984})}\BibitemShut {NoStop}%
\bibitem [{\citenamefont {Nishi}\ \emph {et~al.}(2011)\citenamefont {Nishi},
  \citenamefont {Yamauchi}, \citenamefont {Marukame}, \citenamefont
  {Kinoshita}, \citenamefont {Koga},\ and\ \citenamefont {Kato}}]{Nis11}%
  \BibitemOpen
  \bibfield  {author} {\bibinfo {author} {\bibfnamefont {Y.}~\bibnamefont
  {Nishi}}, \bibinfo {author} {\bibfnamefont {T.}~\bibnamefont {Yamauchi}},
  \bibinfo {author} {\bibfnamefont {T.}~\bibnamefont {Marukame}}, \bibinfo
  {author} {\bibfnamefont {A.}~\bibnamefont {Kinoshita}}, \bibinfo {author}
  {\bibfnamefont {J.}~\bibnamefont {Koga}}, \ and\ \bibinfo {author}
  {\bibfnamefont {K.}~\bibnamefont {Kato}},\ }\href@noop {} {\bibfield
  {journal} {\bibinfo  {journal} {Phys. Rev. B}\ }\textbf {\bibinfo {volume}
  {84}},\ \bibinfo {pages} {115323} (\bibinfo {year} {2011})}\BibitemShut
  {NoStop}%
\bibitem [{\citenamefont {Ehrhardt}\ \emph {et~al.}(2013)\citenamefont
  {Ehrhardt}, \citenamefont {Ferblantier}, \citenamefont {Muller},
  \citenamefont {Ulhaq-Bouillet}, \citenamefont {Rinnert},\ and\ \citenamefont
  {Slaoui}}]{Ehrh13}%
  \BibitemOpen
  \bibfield  {author} {\bibinfo {author} {\bibfnamefont {F.}~\bibnamefont
  {Ehrhardt}}, \bibinfo {author} {\bibfnamefont {G.}~\bibnamefont
  {Ferblantier}}, \bibinfo {author} {\bibfnamefont {D.}~\bibnamefont {Muller}},
  \bibinfo {author} {\bibfnamefont {C.}~\bibnamefont {Ulhaq-Bouillet}},
  \bibinfo {author} {\bibfnamefont {H.}~\bibnamefont {Rinnert}}, \ and\
  \bibinfo {author} {\bibfnamefont {A.}~\bibnamefont {Slaoui}},\ }\href@noop {}
  {\bibfield  {journal} {\bibinfo  {journal} {J. Appl. Phys.}\ }\textbf
  {\bibinfo {volume} {114}},\ \bibinfo {pages} {033528} (\bibinfo {year}
  {2013})}\BibitemShut {NoStop}%
\bibitem [{\citenamefont {Helms}\ and\ \citenamefont
  {Poindexter}(1994)}]{Helm94}%
  \BibitemOpen
  \bibfield  {author} {\bibinfo {author} {\bibfnamefont {C.~R.}\ \bibnamefont
  {Helms}}\ and\ \bibinfo {author} {\bibfnamefont {E.~H.}\ \bibnamefont
  {Poindexter}},\ }\href@noop {} {\bibfield  {journal} {\bibinfo  {journal}
  {Rep. Progr. Phys.}\ }\textbf {\bibinfo {volume} {57}},\ \bibinfo {pages}
  {791} (\bibinfo {year} {1994})}\BibitemShut {NoStop}%
\bibitem [{\citenamefont {Keunen}\ \emph {et~al.}(2011)\citenamefont {Keunen},
  \citenamefont {A},\ and\ \citenamefont {Afanas{\'e}v}}]{Keun11}%
  \BibitemOpen
  \bibfield  {author} {\bibinfo {author} {\bibfnamefont {K.}~\bibnamefont
  {Keunen}}, \bibinfo {author} {\bibfnamefont {A.~S.}\ \bibnamefont {A}}, \
  and\ \bibinfo {author} {\bibfnamefont {V.~V.}\ \bibnamefont {Afanas{\'e}v}},\
  }\href@noop {} {\bibfield  {journal} {\bibinfo  {journal} {Appl. Phys.
  Lett.}\ }\textbf {\bibinfo {volume} {98}},\ \bibinfo {pages} {213503}
  (\bibinfo {year} {2011})}\BibitemShut {NoStop}%
\bibitem [{\citenamefont {Jivanescu}\ \emph {et~al.}(2008)\citenamefont
  {Jivanescu}, \citenamefont {Stesmans},\ and\ \citenamefont
  {Zacharias}}]{Jiva08}%
  \BibitemOpen
  \bibfield  {author} {\bibinfo {author} {\bibfnamefont {M.}~\bibnamefont
  {Jivanescu}}, \bibinfo {author} {\bibfnamefont {A.}~\bibnamefont {Stesmans}},
  \ and\ \bibinfo {author} {\bibfnamefont {M.}~\bibnamefont {Zacharias}},\
  }\href@noop {} {\bibfield  {journal} {\bibinfo  {journal} {J. Appl. Phys.}\
  }\textbf {\bibinfo {volume} {104}},\ \bibinfo {pages} {103518} (\bibinfo
  {year} {2008})}\BibitemShut {NoStop}%
\bibitem [{\citenamefont {Schuppler}\ \emph {et~al.}(1994)\citenamefont
  {Schuppler}, \citenamefont {Friedman}, \citenamefont {Marcus}, \citenamefont
  {Adler}, \citenamefont {Xie}, \citenamefont {Ross}, \citenamefont {Harris},
  \citenamefont {Brown}, \citenamefont {Chabal}, \citenamefont {Brus},\ and\
  \citenamefont {Citrin}}]{Schu94}%
  \BibitemOpen
  \bibfield  {author} {\bibinfo {author} {\bibfnamefont {S.}~\bibnamefont
  {Schuppler}}, \bibinfo {author} {\bibfnamefont {S.~L.}\ \bibnamefont
  {Friedman}}, \bibinfo {author} {\bibfnamefont {M.~A.}\ \bibnamefont
  {Marcus}}, \bibinfo {author} {\bibfnamefont {D.~L.}\ \bibnamefont {Adler}},
  \bibinfo {author} {\bibfnamefont {Y.~H.}\ \bibnamefont {Xie}}, \bibinfo
  {author} {\bibfnamefont {F.~M.}\ \bibnamefont {Ross}}, \bibinfo {author}
  {\bibfnamefont {T.~D.}\ \bibnamefont {Harris}}, \bibinfo {author}
  {\bibfnamefont {W.~L.}\ \bibnamefont {Brown}}, \bibinfo {author}
  {\bibfnamefont {Y.~J.}\ \bibnamefont {Chabal}}, \bibinfo {author}
  {\bibfnamefont {L.~E.}\ \bibnamefont {Brus}}, \ and\ \bibinfo {author}
  {\bibfnamefont {P.~H.}\ \bibnamefont {Citrin}},\ }\href@noop {} {\bibfield
  {journal} {\bibinfo  {journal} {Phys. Rev. Lett.}\ }\textbf {\bibinfo
  {volume} {72}},\ \bibinfo {pages} {2648} (\bibinfo {year}
  {1994})}\BibitemShut {NoStop}%
\bibitem [{\citenamefont {K\"{o}nig}\ \emph {et~al.}(2006)\citenamefont
  {K\"{o}nig}, \citenamefont {Green},\ and\ \citenamefont {Conibeer}}]{Koe06}%
  \BibitemOpen
  \bibfield  {author} {\bibinfo {author} {\bibfnamefont {D.}~\bibnamefont
  {K\"{o}nig}}, \bibinfo {author} {\bibfnamefont {M.~A.}\ \bibnamefont
  {Green}}, \ and\ \bibinfo {author} {\bibfnamefont {G.}~\bibnamefont
  {Conibeer}},\ }in\ \href@noop {} {\emph {\bibinfo {booktitle} {Proc. of
  21$^{\mathrm{st}}$ European Photovoltaics Science and Engineering
  Conference}}}\ (\bibinfo  {publisher} {WIP Munich},\ \bibinfo {address}
  {Dresden, Germany},\ \bibinfo {year} {2006})\ \bibinfo {note} {$\!\!\!$,
  presentation 1CO.6.4, CD-ROM}\BibitemShut {NoStop}%
\bibitem [{\citenamefont {Okada}\ \emph {et~al.}(1984)\citenamefont {Okada},
  \citenamefont {Iwaki}, \citenamefont {Yamamoto}, \citenamefont {Kasahara},\
  and\ \citenamefont {Abe}}]{Okad84}%
  \BibitemOpen
  \bibfield  {author} {\bibinfo {author} {\bibfnamefont {T.}~\bibnamefont
  {Okada}}, \bibinfo {author} {\bibfnamefont {T.}~\bibnamefont {Iwaki}},
  \bibinfo {author} {\bibfnamefont {K.}~\bibnamefont {Yamamoto}}, \bibinfo
  {author} {\bibfnamefont {H.}~\bibnamefont {Kasahara}}, \ and\ \bibinfo
  {author} {\bibfnamefont {K.}~\bibnamefont {Abe}},\ }\href@noop {} {\bibfield
  {journal} {\bibinfo  {journal} {Sol. Stat. Comm.}\ }\textbf {\bibinfo
  {volume} {49}},\ \bibinfo {pages} {809} (\bibinfo {year} {1984})}\BibitemShut
  {NoStop}%
\bibitem [{\citenamefont {K\r{u}sov\`{a}}\ \emph {et~al.}(2012)\citenamefont
  {K\r{u}sov\`{a}}, \citenamefont {Ondi\v{c}}, \citenamefont
  {Klime\v{s}ov\`{a}}, \citenamefont {Herynkov\`{a}}, \citenamefont {Pelant},
  \citenamefont {Dani\v{s}}, \citenamefont {Valenta}, \citenamefont {Gallart},
  \citenamefont {Ziegler}, \citenamefont {H{\"o}nerlage},\ and\ \citenamefont
  {Gilliot}}]{Kuso12}%
  \BibitemOpen
  \bibfield  {author} {\bibinfo {author} {\bibfnamefont {K.}~\bibnamefont
  {K\r{u}sov\`{a}}}, \bibinfo {author} {\bibfnamefont {L.}~\bibnamefont
  {Ondi\v{c}}}, \bibinfo {author} {\bibfnamefont {E.}~\bibnamefont
  {Klime\v{s}ov\`{a}}}, \bibinfo {author} {\bibfnamefont {K.}~\bibnamefont
  {Herynkov\`{a}}}, \bibinfo {author} {\bibfnamefont {I.}~\bibnamefont
  {Pelant}}, \bibinfo {author} {\bibfnamefont {S.}~\bibnamefont {Dani\v{s}}},
  \bibinfo {author} {\bibfnamefont {J.}~\bibnamefont {Valenta}}, \bibinfo
  {author} {\bibfnamefont {M.}~\bibnamefont {Gallart}}, \bibinfo {author}
  {\bibfnamefont {M.}~\bibnamefont {Ziegler}}, \bibinfo {author} {\bibfnamefont
  {B.}~\bibnamefont {H{\"o}nerlage}}, \ and\ \bibinfo {author} {\bibfnamefont
  {P.}~\bibnamefont {Gilliot}},\ }\href@noop {} {\bibfield  {journal} {\bibinfo
   {journal} {Appl. Phys. Lett.}\ }\textbf {\bibinfo {volume} {101}},\ \bibinfo
  {pages} {143101} (\bibinfo {year} {2012})}\BibitemShut {NoStop}%
\bibitem [{\citenamefont {Valenta}\ \emph {et~al.}(2002)\citenamefont
  {Valenta}, \citenamefont {Juhasz},\ and\ \citenamefont {Linnros}}]{Vale02}%
  \BibitemOpen
  \bibfield  {author} {\bibinfo {author} {\bibfnamefont {J.}~\bibnamefont
  {Valenta}}, \bibinfo {author} {\bibfnamefont {R.}~\bibnamefont {Juhasz}}, \
  and\ \bibinfo {author} {\bibfnamefont {J.}~\bibnamefont {Linnros}},\
  }\href@noop {} {\bibfield  {journal} {\bibinfo  {journal} {Appl. Phys.
  Lett.}\ }\textbf {\bibinfo {volume} {80}},\ \bibinfo {pages} {1070} (\bibinfo
  {year} {2002})}\BibitemShut {NoStop}%
\bibitem [{\citenamefont {Valenta}\ \emph {et~al.}(2004)\citenamefont
  {Valenta}, \citenamefont {Lalic},\ and\ \citenamefont {Linnros}}]{Vale04}%
  \BibitemOpen
  \bibfield  {author} {\bibinfo {author} {\bibfnamefont {J.}~\bibnamefont
  {Valenta}}, \bibinfo {author} {\bibfnamefont {N.}~\bibnamefont {Lalic}}, \
  and\ \bibinfo {author} {\bibfnamefont {J.}~\bibnamefont {Linnros}},\
  }\href@noop {} {\bibfield  {journal} {\bibinfo  {journal} {Appl. Phys.
  Lett.}\ }\textbf {\bibinfo {volume} {84}},\ \bibinfo {pages} {1459} (\bibinfo
  {year} {2004})}\BibitemShut {NoStop}%
\bibitem [{\citenamefont {Ibanez}\ \emph {et~al.}(2015)\citenamefont {Ibanez},
  \citenamefont {Hernandez}, \citenamefont {Lopez-Vidrier}, \citenamefont
  {Hiller}, \citenamefont {Gutsch}, \citenamefont {Zacharias}, \citenamefont
  {Segura}, \citenamefont {Valenta},\ and\ \citenamefont {Garrido}}]{Iban15}%
  \BibitemOpen
  \bibfield  {author} {\bibinfo {author} {\bibfnamefont {J.}~\bibnamefont
  {Ibanez}}, \bibinfo {author} {\bibfnamefont {S.}~\bibnamefont {Hernandez}},
  \bibinfo {author} {\bibfnamefont {J.}~\bibnamefont {Lopez-Vidrier}}, \bibinfo
  {author} {\bibfnamefont {D.}~\bibnamefont {Hiller}}, \bibinfo {author}
  {\bibfnamefont {S.}~\bibnamefont {Gutsch}}, \bibinfo {author} {\bibfnamefont
  {M.}~\bibnamefont {Zacharias}}, \bibinfo {author} {\bibfnamefont
  {A.}~\bibnamefont {Segura}}, \bibinfo {author} {\bibfnamefont
  {J.}~\bibnamefont {Valenta}}, \ and\ \bibinfo {author} {\bibfnamefont
  {B.}~\bibnamefont {Garrido}},\ }\href@noop {} {\bibfield  {journal} {\bibinfo
   {journal} {Phys. Rev. B}\ }\textbf {\bibinfo {volume} {92}},\ \bibinfo
  {pages} {035432} (\bibinfo {year} {2015})}\BibitemShut {NoStop}%
\bibitem [{\citenamefont {Heitmann}\ \emph {et~al.}(2005)\citenamefont
  {Heitmann}, \citenamefont {M\"{u}ller}, \citenamefont {Zacharias},\ and\
  \citenamefont {G\"{o}sele}}]{Heit05}%
  \BibitemOpen
  \bibfield  {author} {\bibinfo {author} {\bibfnamefont {J.}~\bibnamefont
  {Heitmann}}, \bibinfo {author} {\bibfnamefont {F.}~\bibnamefont
  {M\"{u}ller}}, \bibinfo {author} {\bibfnamefont {M.}~\bibnamefont
  {Zacharias}}, \ and\ \bibinfo {author} {\bibfnamefont {U.}~\bibnamefont
  {G\"{o}sele}},\ }\href@noop {} {\bibfield  {journal} {\bibinfo  {journal}
  {Adv. Mater.}\ }\textbf {\bibinfo {volume} {17}},\ \bibinfo {pages} {795}
  (\bibinfo {year} {2005})}\BibitemShut {NoStop}%
\end{thebibliography}%

\end{document}